\documentclass[a4paper,12pt,final]{article}
\usepackage{etex}

\usepackage[sc]{mathpazo}

\usepackage[table]{xcolor}

\usepackage{threeparttable}
\usepackage{standalone}
\usepackage{booktabs, dcolumn}
\usepackage{booktabs}

\usepackage{afterpage}

\usepackage{rotating}
\usepackage{tikz}

\usepackage{latexsym}
\usepackage{amsfonts}
\usepackage{amssymb}
\usepackage{amsthm}
\usepackage{amsmath}
\usepackage[mathscr]{eucal}
\usepackage{multicol}
\usepackage{multirow}
\usepackage{setspace,graphicx,amsmath,amsfonts,amssymb,amsthm}
\usepackage{pdfpages}
\usepackage{pdflscape}

\usepackage{anyfontsize}
\usepackage{t1enc}

\usepackage{caption}
\usepackage{subcaption}
\DeclareCaptionFont{captiofont}{\fontsize{9pt}{10.8pt}\selectfont}
\usepackage{verbatim}
\usepackage[plainpages=false,
breaklinks=true,
colorlinks=true,
urlcolor=gray,
citecolor=blue,
linkcolor=red,
bookmarks=true,
bookmarksopen=true,
bookmarksopenlevel=1,
pdfstartview=FitH,
pdfview=FitH]{hyperref}

\usepackage{color,hyperref}

\usepackage[round]{natbib}

\usepackage[lmargin=1in, rmargin=1in, tmargin=1in, bmargin=1in, paperwidth=220mm, paperheight=290mm]{geometry}

\usepackage{booktabs}
\usepackage{tabulary}
\usepackage{tabularx}
\usepackage{multirow}
\usepackage{array}

\usepackage{enumerate}
\usepackage{sectsty}
\usepackage{lipsum}
\sectionfont{\centering}

\usepackage{tocloft}
\usepackage{titlesec}
\titleformat{\section}
{\large\bfseries}{\thesection}{0.5em}{}

\renewcommand\thesection{\arabic{section}}

\renewcommand\thesubsection{\thesection.\arabic{subsection}}
\titleformat{\subsection}
{\normalfont\itshape}{\thesubsection}{0.5em}{}

\renewcommand\thesubsubsection{\thesection.\arabic{subsection}.\arabic{subsubsection}}
\titleformat{\subsubsection}{\normalfont\itshape}{\thesubsubsection}{0.5em}{}

\usepackage{setspace}

\titlespacing\section{0pt}{10pt plus 4pt minus 2pt}{5pt plus 2pt minus 2pt}
\titlespacing\subsection{0pt}{10pt plus 4pt minus 2pt}{0pt plus 2pt minus 2pt}
\titlespacing\subsubsection{0pt}{10pt plus 4pt minus 2pt}{0pt plus 2pt minus 2pt}

\providecommand{\keywords}[1]{\textbf{Keywords:}  #1}
\providecommand{\JEL}[1]{\textbf{JEL:}  #1}

\makeatletter
\newcommand*{\myfnsymbolsingle}[1]{%
\ensuremath{%
\ifcase#1
\or 
*%
\or 
\dagger
\or 
\ddagger
\or 
\mathsection
\or 
\mathparagraph
\else 
\@ctrerr
\fi
}%
}
\makeatother

\usepackage{alphalph}
\newalphalph{\myfnsymbolmult}[mult]{\myfnsymbolsingle}{}

\renewcommand*{\thefootnote}{%
\myfnsymbolmult{\value{footnote}}%
}

\makeatletter
\def\@xfootnote[#1]{%
\protected@xdef\@thefnmark{#1}%
\@footnotemark\@footnotetext}
\makeatother

\usepackage{bbding}         

\usepackage[symbol]{footmisc}

\usepackage{chngcntr}

\usepackage{scrwfile}
\TOCclone[\contentsname]{toc}{atoc}

\AfterTOCHead[toc]{%
}
\AfterTOCHead[atoc]{%
\edef\maintocdepth{\the\value{tocdepth}}%
\value{tocdepth}=-10000\relax%
}

\usepackage{blindtext}

\definecolor{myred}{rgb}{0.89412, 0.10196, 0.10980}
\definecolor{myblue}{rgb}{0.21569, 0.49412, 0.72157}
\definecolor{mygreen}{rgb}{0.30196, 0.68627, 0.29020}
\definecolor{mygray}{rgb}{0.90, 0.90, 0.90}

\begin{document}

\setcounter{footnote}{0}

\newpage
\hypersetup{colorlinks,linkcolor=black} 

\linespread{1}



\title{\Large \setcounter{footnote}{0}Learning Probability Distributions \\ in Macroeconomics and Finance\thanks{We are grateful to Wolfgang Hardle, Lukas Vacha, Martin Hronec, Frantisek Cech, and the participants at various conferences and research seminars for many useful comments, suggestions, and discussions. We gratefully acknowledge the support from the Czech Science Foundation under the EXPRO GX19-28231X project. We provide the computational package \texttt{DistrNN.jl} in \textsf{JULIA} available at \url{https://github.com/barunik/DistrNN.jl} that allows one to obtain our measures on data the researcher desires. }
\vspace{20pt}}

\author{\setcounter{footnote}{0}Jozef Barun\'{i}k\thanks{Institute of Economic Studies, Charles University, Opletalova 26, 110 00, Prague, CR and Institute of Information Theory and Automation, Czech Academy of Sciences , Pod Vodarenskou Vezi 4, 18200, Prague, Czech Republic. E-mail: \url{barunik@utia.cas.cz }\,\, Web: \href{https://barunik.github.io/}{barunik.github.io}} \\ {\small\textit{Charles University and}} \\
{\small\textit{Czech Academy of Sciences}}
\and
\and
\setcounter{footnote}{6}Lubo\v{s} Hanus\thanks{Institute of Economic Studies, Charles University, Opletalova 26, 110 00, Prague, CR and Institute of Information Theory and Automation, Academy of Sciences of the Czech Republic, Pod Vodarenskou Vezi 4, 18200, Prague, Czech Republic. E-mail: \url{hanusl@utia.cas.cz}} \\
{\small\textit{Charles University and}} \\
{\small\textit{Czech Academy of Sciences}}}

\date{\normalsize\vspace{2em} \hspace{2em} First draft: June 2019 \hspace{4em} This draft: \today }

\maketitle

\linespread{1.0}

\begin{abstract}
We propose a deep learning approach to probabilistic forecasting of macroeconomic and financial time series. Being able to learn complex patterns from a data rich environment, our approach is useful for a decision making that depends on uncertainty of large number of economic outcomes. Specifically, it is informative to agents facing asymmetric dependence of their loss on outcomes from possibly non-Gaussian and non-linear variables. We show the usefulness of the proposed approach on the two distinct datasets where a machine learns the pattern from data. First, we construct macroeconomic fan charts that reflect information from high-dimensional data set. Second, we illustrate gains in prediction of stock return distributions which are heavy tailed, asymmetric and suffer from low signal-to-noise ratio. 
\vspace{10pt}

\keywords{Distributional forecasting, machine learning, deep learning, probability, economic time series}

\JEL{C45, C53, E17, E37}
\end{abstract}

\renewcommand{\thefootnote}{\arabic{footnote}}
\setcounter{footnote}{0}

\newpage
\tableofcontents

\newpage
\hypersetup{colorlinks,linkcolor=red} 
\section{Introduction}
\label{sec: intro}

Despite advances in data availability, theory, and computational power, economics have not enjoyed dramatic improvements in forecast accuracy of economic variables over past decades \citep{stock2017twenty}. A fundamental problem underlying the lack of success is that economic variables are difficult to forecast at its nature. Economic forecasters whose decisions depend on such uncertainty hence need to focus on communicating full predictive distribution of the variable surrounding point estimates.\footnote{The Bank of England was an early leader recognizing this need and started to communicate the uncertainty as fan charts to the public.} At the same time, economists being keen to use large number of series to understand fluctuations in economic data  collect data unimaginable decades ago.\footnote{Already a century ago in 1920s, Harvard Economic Service provided economic indexes and forecasts based on available data \citep{friedman2009harvard}. Later, 1277 time series were used to study business cycles by \cite{lerner1947measuring}.} With the explosive growth in data volume, velocity, and variety, necessity to unlock the information hidden in big data is becoming a key theme in economics \citep{diebold2020big}. Challenged by proliferation of parameters, heavy critique of arbitrarily chosen restrictions on reduced as well as structured models in past decades, economists wishing to explore possibly rich information content of new datasets recently turned their hopes 
towards machine learning \citep{mullainathan2017machine}. In this paper, we explore use of machine learning for information-rich uncertainty forecasts. We develop a  distributional machine learning methods based on deep learning and recurrent network techniques to provide probabilistic forecasts that reflect time series dynamics of possibly large amounts of available information. Such data-driven probabilistic forecasts aim to improve current state in forecasting and communicating uncertainty in economics.

Uncertainty surrounding any step in decision-making is key to understand for financial operations, central and retail banking as well as researchers and practitioners trying to minimize the risk of their decisions, appropriate plans and assisting in the design and implementation of economic policies. Yet even after decades of research, a conditional mean forecast often serves economists as a convenient tool for measuring the central tendency of a target variable, or simply as a best guess about future outcomes of a variable. Variance forecast accompanying it often serves as best expectation about uncertainty and future risk. Such predictions are nevertheless not fully informative in case a decision maker is facing asymmetric dependence of her loss on outcomes from possibly non-Gaussian variables. With an uncertainty being key ingredient in economic decision making, shift to probabilistic forecasting also shifts our hopes towards obtaining better expectations about entire distributions of economic variables. A nontrivial question is how do we make such forecasts, especially utilizing available data.

Traditionally, distribution forecasts are made using time-series models, surveys, or are collected in real-time.\footnote{Methods for constructing distribution forecasts are reviewed in a special issues on ``Density Forecasting in Economics and Finance'' \citep{timmermann2000density} and ``Probability Forecasting'' \citep{gneiting2008probabilistic} for collection of papers.} With rapid improvements in accessibility and availability of large datasets we believe one can improve description of uncertainty substantially utilizing methods that focus on learning patterns from data. In line with recent endeavors of economists to move away from exclusive dependence on models towards machine learning approaches \citep{athey2019machine} when it makes sense to utilize data and improve our understanding of the problem \citep{mullainathan2017machine}, we propose to use deep machine learning to learn the complex patterns in data and return the user a prediction of a full distribution. 

A key idea of (machine) learning that can be thought of as inferring plausible models to explain observed data recently attracted number of researchers who document how learning patterns from data can be useful.\footnote{\citet{mullainathan2017machine,sirignano2016deep,gu2020empirical,heaton2017deep,tobek2020does,bianchi2020bond,israel2020can,iworiso2020directional,feng2018deep,coulombe2020machine}} Surge in the literature and increasing number of applications in economics focus mostly on cross-sectional data and ultimately on point forecasts. While machine can use such models to make predictions about future data, uncertainty plays fundamental part. At the same time, data being key ingredients of all machine-learning systems are useless on their own until one extracts knowledge or inferences from them. Shifting focus from point forecasts towards probabilistic forecasting using big data is essential next step for economists wishing to explore what computer science has to offer.

We contribute to this debate by exploring machine learning in time series context and developing a machine learning strategy to forecast the full distributions in possibly large dimensional setting. We argue that deep learning in combination with recurrent neural network offers useful tool for distribution prediction without the need of model specification, simply learning the distributions from data. While ability to outperform alternative methods on specific data sets in terms of out-of-sample predictive power is valuable in practice, such performance is rarely explicitly acknowledged as a goal to be addressed in econometrics. As \cite{mullainathan2017machine} highlights, some substantive problems are naturally cast as prediction problems, and assessing their goodness of fit on a test set may be sufficient for purpose of the analysis. We believe that distribution prediction task in data-rich environment is one of such important problems in economics where machine learning could be helpful for a researcher, policy maker or a practitioner. 

What are the challenges specific to probabilistic forecasting of economic variables? Time series as stock returns, electricity prices,  traffic data, or macroeconomic series display distributions that can not be captured by convenient Gaussian distribution and hence are not fully characterized by means and variances. These distributions show heavy tails, they are asymmetric, and often violate stationarity. Further, the data contain irregularities, difficult to predict spikes, and regime shifts. Hence complete information about probability of future outcomes given the past information that can be mapped into different representations to construct prediction intervals or probability distribution functions reflecting the data are needed. Such fully approximated distribution function provides a comprehensive information about uncertainty of future observations. 

Vast majority of studies that focus on the prediction of conditional return distributions characterize the cumulative conditional distribution by a collection of conditional quantiles \citep{engle2004caviar,vzikevs2014semi}. In contrast, \cite{leorato2015comparing} argue that collection of conditional probabilities that describe the cumulative distribution function using set of separate logistic regressions \citep{foresi1995conditional} provide better approach. Following decades resulted in few contributions exploring distributional regressions \citep{chernozhukov2013inference,fortin2011decomposition,rothe2012partial} including attempts to overcome problem of monotonicity of the forecasts \citep{barunik2019simple} using ordered logistic parametrization. Another important strand of literature focuses on Bayesian forecasting where uncertainty is characterized by probabilities automatically \citep{geweke2006bayesian,lahiri2010bayesian}.
At the same time, literature in computer science attempts to use machine learning in prediction of distributions. 
These attempts are similar to traditional methods and mostly rely on approximation of some pre-specified distribution, such as first two moments.\footnote{\citet{duan2019ngboost} applies the natural gradient boosting algorithm to estimate parameters for conditional probability distribution, while assuming homoskedasticity. \citet{salinas2020deepar} build an autoregressive recurrent neural network, which learns mean and standard deviation for Gaussian, and mean and shape parameter for Negative binomial. \citet{lim2020deep} classifies price movements for high-frequency trading via deep probabilistic modelling when optimizing parameters of different families of distribution. Although similarly to \citet{salinas2020deepar}, \citet{chen2020probabilistic} proposes to use a deep temporal convolutional neural network to estimate parameters of Gaussian distribution to model probabilistic forecast, and they further propose to use the same architecture for non-parametric estimation of quantile regression. The second approach is distribution-free and can produce more robust results. Another study forecasts distributions via direct quantiles using recurrent neural network, \citet{wen2017multi} also perform a multi-horizon predictions. Quantile function represented by spline combined with recurrent neural network proposed by \citet{gasthaus2019probabilistic} is a distribution-free approach with objective function based on CRPS score \citep{gneiting2007strictly} constructed with respect to monotonicity of quantile function.  \citet{hu2019distribution} build deep neural networks to obtain distribution-free probability distribution where one of the steps in the procedure is to obtain cumulative distribution estimates. \citet{januschowski2020criteria} provide a detailed discussion about ML methods for forecasting. The text discusses way of distinction between "statistical" and "ML" methods adapted in time.} To the best of our knowledge, literature have not yet moved to fully non-parametric approaches approximating the data structures in the context of distributional forecasting in economics and finance.

Why we should believe that machine learning can improve probability forecasts? Classical time series econometrics \citep{box2015time,hyndman2008forecasting} mainly focuses on predetermined autocorrelation or seasonality structures in data that are parametrized. With large amount of time series available to researchers, these methods quickly become infeasible and unable to explore more complex data structures. Keeping in mind a famous wisdom that \textit{``all models are wrong..., but some of them are useful.''} \citep{box1987empirical}, modern machine learning methods are able to easily overcome these problems. Being a powerful tool for approximation of a complex and unknown data structures \citep{kuan1994artificial}, these methods can be useful in number of application problems where data contain rich information structure and we can not describe it satisfactorily by a simplifying model. Overcoming the longstanding problem of computational intensity of such data-driven approach with advances in computer sciences adds to the temptation in using these methods for addressing new problems such as distribution predictions. 

Our main contribution to the literature is that we propose how to use deep learning techniques as an useful tool for approximation and prediction of conditional distributions in data-rich environment. Our distributional neural network combines deep learning with recurrent neural networks, it is capable of predicting an entire distribution of a time series and allows to use large amounts of variables. We frame our approach as a multi-output neural network, which returns approximate probability functions of the distribution. The novelty of our approach hence lies in learning of the entire conditional distribution using (deep) recurrent networks from big data. The proposed network is also capable to capture time-variation of distributions when, for example, dealing with highly dynamic data and recover longer and more complex time dependence structures present in data. Our framework generalizes binary choice models \citep{foresi1995conditional,barunik2019simple} and together with the state-of-the-art machine learning tools forms a new toolkit for economists interested to describe future uncertainty of economic variables. An important contribution is also our novel approach to construction of objective function that fulfills monotonicity of distributional forecasts by introducing penalty function for divergence from monotone behavior.

Two distinct and important economic datasets illustrate how the machine learning approach to probabilistic forecasting may help a decision maker facing uncertainty. First, we use deep learning to construct data-driven macroeconomic fan charts reflecting information contained in large number of variables. Such data-rich fan charts are first of its kind to reflect high-dimensional information of 216 relevant variables and are of great importance for policy makers as they reflect the structures in data and are not influenced by choice of the model. A forecasting model is in contrast  learned from data. Such data-rich fan charts moreover can not be obtained with traditional methods. Second, we study the set of most liquid U.S. stock returns that display asymmetric, heavy-tailed dynamically evolving distributions that are hard to predict due to very low signal to noise ratio.

\section{A Route Towards Probabilistic Forecasting via Deep Learning}
\label{sec: methods}

Let us consider an economic time series $y_t$ collected over $t=1\ldots,T$. The main objective is to approximate the conditional cumulative distribution function $F(y_{t+h}|\mathcal{I}_{t})$ as precisely as possible, and use it for $h$-step-ahead probabilistic forecast made at time $t$ with information $\mathcal{I}_{t}$ containing past values of $y_t$ as well as, possibly, past values of other exogenous observable variables. 

Consider a partition of the support of $y_t$ by $p>1$ fixed thresholds corresponding to set of empirical $\alpha_j$-quantiles $\{q^{\alpha_j}\}_{j=1}^{p}$ where $0<\alpha_1<\alpha_2<\ldots<\alpha_p<1$ are $p$ regularly spaced probability levels on a unit interval $[0,1]$. These partitions are further time-varying, thus in general the elements of the partition are indexed implicitly by $t$.

The main goal then is to approximate a collection of conditional probabilities corresponding to the empirical quantiles such as 
$$\Big\{F(q^{\alpha_1}),\ldots,F(q^{\alpha_p})\Big\}=\Big\{ \Pr\Big(y_{t+h}\le q^{\alpha_1}|\mathcal{I}_{t}\Big),\ldots,\Pr\Big(y_{t+h}\le q^{\alpha_p}|\mathcal{I}_t\Big) \Big\}$$
for the collection of thresholds $1,\ldots,p$. One convenient way of estimating such quantities is distributional regression. \cite{foresi1995conditional} noted that several binary regressions serve as a good partial description of the conditional distribution. To estimate conditional distribution, one can simply consider distribution regression model
\begin{equation}
\label{eq:probit}
\Pr ( y_{t+h} \leq q^{\alpha_j} | \mathcal{I}_{t} ) = \Lambda(\beta_{j}),
\end{equation}
where $\Lambda : z \to [0,1]$ is a known (monotonically increasing) link function, such as logit, probit, linear, log-log functions\footnote{As discussed by \cite{chernozhukov2013inference}, log-log link nests the Cox model making distribution regression important. } and $\beta(.)$ is an unknown function-valued parameter to be determined. In contrast to estimating separate models for separate thresholds, \cite{chernozhukov2013inference} considered continuum of binary regressions, and argued it provides a coherent and flexible model for the entire conditional distribution as well as useful alternative to \cite{koenker1978regression}'s quantile regression. Alternatively, \cite{barunik2019simple} propose to tie the coefficients of predictors in an ordered logit model via smooth dependence on corresponding probability levels. While being able to forecast entire distribution and keeping $0 < F_j < 1$ and  $0 < F_1(.) < F_2(.) < \dots F_p(.) < 1$, the approach still depends on heavy parametrization suited for a specific problem of the time series considered making it an infeasible approach for larger number of variables.

\subsection{(Deep) Machine Learning}

Such probabilistic forecasts heavily depend on the model parametrization and with growing number of covariates become quickly infeasible. Stationarity of data at hand is also requirement that complicates forecasts as it is hard to achieve in many cases. In sharp contrast to such approach, we propose more flexible and general way to the distribution regression via deep learning. We propose a novel multiple output neural network we refer to as a distribution neural network (DistrNN). Our approach aims to uncover non-linear and mostly complex relationship of time series without specifying strict parametric structure and without requiring strict assumptions about data, while focusing on the out-of-sample predictive power of the model. 

Machine learning has a long history in economics and finance \citep*{hutchinson1994nonparametric,kuan1994artificial,racine2001nonlinear,baillie2007testing}. At its core, one may perceive machine learning as a general statistical analysis that economists can use to capture complex relationships that are hidden when using simple linear methods. As emphasized by \cite{breiman2001statistical}, maximizing prediction accuracy in the face of an unknown model differentiates machine learning from the more traditional statistical objective of estimating a model assuming a data generating process. Building on this, machine learning seeks to choose the most preferable model from an unknown pool of models using innovative optimization techniques. As opposed to traditional measures of fit, machine learning focuses on the out-of-sample forecasting performance and understanding the bias-variance trade-off; as well as using data driven techniques that concentrate on finding structures in large datasets. Further, if one dismisses the ``black-box'' view of machine learning as a misconception \citep*{lopez2019beyond}, it seems nothing should stop a researcher from exploring the power of these methods to solve problems like probabilistic forecasting. However the problems in economics differ from a typical machine learning applications in many aspects. In order to enjoy the benefits of machine learning, a user needs to understand key challenges brought by data.\footnote{For example \cite{israel2020can} note that machine learning applied to finance is challenged by small sample sizes, naturally low signal-to-noise ratios making market behavior difficult to predict and the dynamic character of markets. Because of these critical issues, the benefits of machine learning are not so obvious as in other fields and research into understanding how impactful machine learning can be for asset management is just emerging. With the surge in deep learning literature, machine learning applications in finance have begun to emerge \citep*{heaton2017deep,feng2018deep,bryzgalova2019forest,bianchi2020bond,chen2020deep,gu2020empirical,tobek2020does,zhang2020deep}.}

Deep feedforward networks, also often called feedforward neural networks, or multilayer perceptrons lie at heart of deep learning models and are universal approximators that can learn any functional relationship between input and output variables with sufficient data \cite{kuan1994artificial}. As a class of supervised learning methods, these approaches are used for classification, recognition and prediction. While being increasingly popular in economics for solution of particular problems \cite{athey2019machine,mullainathan2017machine}, probabilistic forecasting have not been explored by the literature yet.

This motivates us to reformulate distribution regression into a more general and flexible distributional neural network. The functional form of the new network is driven by data and we may relax assumptions on the distribution of the data, parametric model as well stationarity of data. The proposed distributional neural network is, as feed-forward network, a hierarchical chain of layers that represents high-dimensional and/or non-linear input variables with the aim to predict the target output variable. Importantly, we approximate the conditional distribution function with multiple outputs of the network as set of probabilities jointly. 

As a first step, we exchange a known link function from Eq. \ref{eq:probit} for an unknown general function $\mathfrak{g}$ that will be approximated by a neural network:
\begin{equation}
    \Pr\left(y_{t+h} \leq q^{\alpha_j} | \mathcal{I}_{t}\right) = \mathfrak{g}_j(\cdot).
    \label{eq:probs_g}
\end{equation}
Next, we consider a set of probabilities corresponding to $0<\alpha_1<\alpha_2<\ldots<\alpha_p<1$ being $p$ regularly spaced levels that characterize conditional distribution function using set of predictors $z_t = (y_{t},x_t^1, ..., x_t^n)^{\top}$, and model them jointly as
\begin{equation}
\Big\{ \Pr\Big(y_{t+h}\le q^{\alpha_1}|z_{t}\Big),\ldots,\Pr\Big(y_{t+h}\le q^{\alpha_p}|z_t\Big) \Big\} = \mathfrak{g}_{W,b}(z_t),
\label{eq: ffnet}
\end{equation}
where $\mathfrak{g}_{W,b}$ is a multiple output neural network with $L$ hidden layers that we name as distributional neural network:
\begin{equation}
 \mathfrak{g}_{W,b}(z_t) = g^{(L)}_{W^{(L)},b^{(L)}} \circ \ldots \circ g^{(1)}_{W^{(1)},b^{(1)}} \left(z_t\right),
\label{eq: ffnetDistr}
\end{equation}
where $W=\left(W^{(1)},\ldots,W^{(L)}\right)$ and $b=\left(b^{(1)},\ldots,b^{(L)}\right)$ are weight matrices and bias vector. Any weight matrix $W^{(\ell)} \in \mathbb{R}^{m\times n}$ contain $m$ neurons as $n$ column vectors $W^{(\ell)} = [w_{\cdot,1}^{(\ell)},\ldots,w_{\cdot,n}^{(\ell)}]$, and $b^{(\ell)}$ are thresholds or activation levels which contribute to the output of a hidden layer allowing the function to be shifted. 

\begin{figure}[!ht]
    \centering\includegraphics[width=\textwidth]{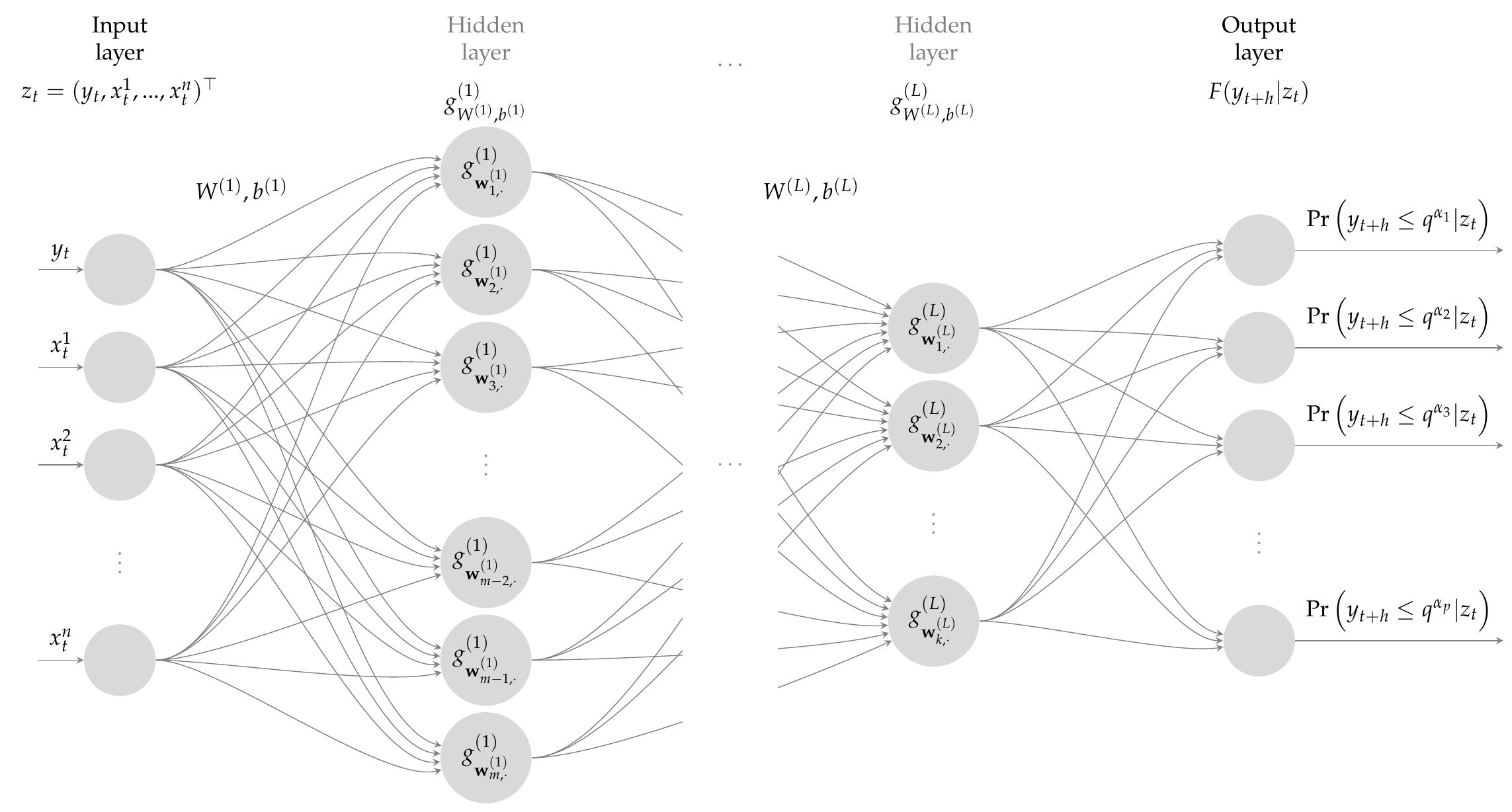}
    \caption{Distributional (Deep) Feed-forward Network.} 
    \vspace{\medskipamount}
        \begin{minipage}{\textwidth} 
            \footnotesize
            An illustration of a multiple output (deep) neural network $\mathfrak{g}_{W,b}(z_t)$ to model the collection of conditional probabilities $\Big\{ \Pr\Big(y_{t+h}\le q^{\alpha_1}|z_{t}\Big),\ldots,\Pr\Big(y_{t+h}\le q^{\alpha_p}|z_t\Big) \Big\} $ with set of predictor variables $z_t = (y_{t},x_t^1, ..., x_t^n)^{\top}$. With large number of hidden layers $L$ the network is deep.
        \end{minipage}
    \label{fig: deepnet}
\end{figure}

It is important to note that in sharp contrast to the literature, we consider a multiple output (deep) neural network to characterize collection of probabilities. Before discussing the details of estimation that allow us to keep monotonicity of probabilities, we illustrate the framework. Figure ~\ref{fig: deepnet} illustrates how $l \in {1,...,L}$ hidden layers transform input data into a chain using collection of non-linear activation functions $g^{(1)},\ldots,g^{(L)}$. A commonly used activation functions, $g^{(\ell)}_{W^{(\ell)},b^{(\ell)}}$, used as
\begin{equation*}
g^{(\ell)}_{W^{(\ell)},b^{(\ell)}} := g_{\ell}\left( W^{(\ell)}z_t + b^{(\ell)} \right) = g_{\ell}\left( \sum_{i=1}^{m} W^{(\ell)}_i z_t + b_i^{(\ell)} \right)
\label{eq: ffneuron}
\end{equation*}
are a sigmoid $g_{\ell}(u) = \sigma(u) = 1/(1+\exp(-u))$, rectified linear units $g_{\ell}(u) = \max\{u,0\}$, or $g_{\ell}(u) = \tanh(u)$. 
In case $\mathfrak{g}_{W,b}(z_t)$ is non-linear, neural network complexity grows with increasing number of neurons $m$, and with increasing number of hidden layers $L$ and we build a deep neural network. We use activation function $g^{(L)}(\cdot) = \sigma(\cdot)$ to transform outputs to probabilities. Note that for $L=1$, neural network becomes a simple logistic regression.

\subsection{(Deep) Recurrent Neural Networks}

Predictors used by economists often evolve over time, and hence traditional neural networks assuming independence of data may not approximate relationships sufficiently well. Instead, a Recurrent Neural Network (RNN) that takes into account time series behavior may help in the prediction task. Taking into account sequential nature of data that evolve over time and possess an autocorrelation structure, RNNs are more suitable for many economic problems. In contrast to plain neural networks, hidden layers in recurrent networks are being  updated in a recurrence for every time step of  the sequence meaning that the weights of the network are shared over the sequential data, and hidden states remember the time structure.

\begin{figure}[ht]
    \centering\includegraphics[width=\textwidth]{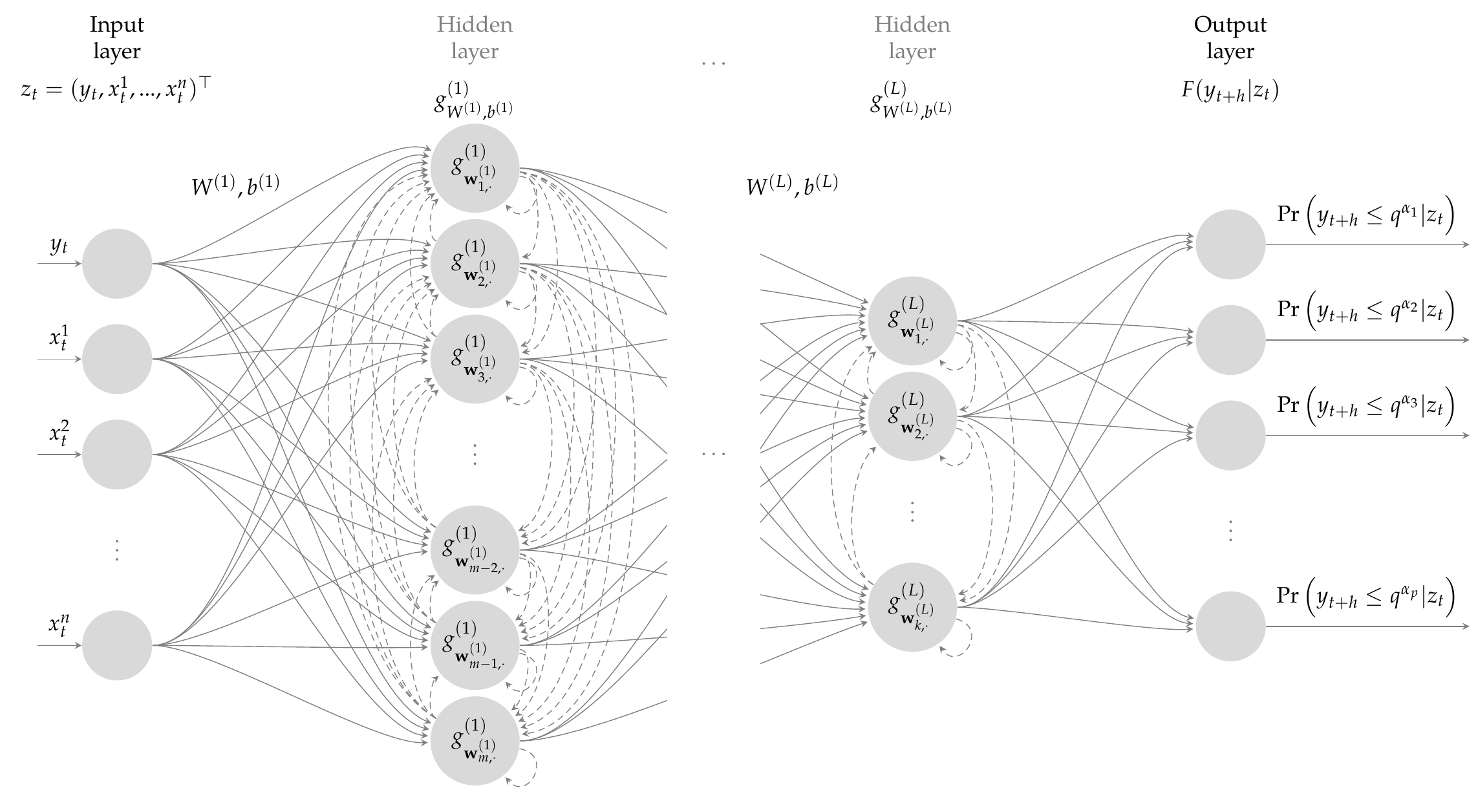}
    \caption{Distributional (Deep) Recurrent Network.}
    \vspace{\medskipamount}
        \begin{minipage}{\textwidth} 
            \footnotesize
            An illustration of a deep recurrent neural network $\mathfrak{g}_{W,b}(z_t)$ that captures relationship between all nodes (solid) and recurrent paths (dashed) in the network at time $t$ to model the collection of conditional probabilities $\Big\{ \Pr\Big(y_{t+h}\le q^{\alpha_1}|z_{t}\Big),\ldots,\Pr\Big(y_{t+h}\le q^{\alpha_p}|z_t\Big) \Big\} $ with set of predictor variables $z_t = (y_{t},x_t^1, ..., x_t^n)^{\top}$. With large number of hidden layers $L$ the network is deep.
        \end{minipage} 
    \label{fig: RNnet}
\end{figure}

Formally, RNNs transform a sequence of input variables to another output sequence with lagged (memory) hidden states
\begin{equation}
h_t = g(W_h h_{t-1} + W_z z_t + b_0). 
\label{eq: hidden}
\end{equation}
Figure~\ref{fig: RNnet} illustrates distinctions of weights where dashed lines correspond to $W_h$ and solid lines to $W_z$. 
Intuitively, RNN is a non-linear generalization of an autoregressive process where lagged variables are transformations of the observed variables. Nevertheless, the structure is only useful when the immediate past is relevant. In case the dynamics are driven by events that are further back in the past, the nodes of the network require even more complex structure.

\subsubsection{Long Short-term Memory (LSTM)} 

As a particular form of recurrent networks, an LSTM provides a solution to the short memory problem by incorporating memory units into the structure \citep{hochreiter1997long} and capture potentially long time dynamics in the time series. Memory units allow the network to learn when to forget previous hidden states and when to update hidden states given new information. Specifically, LSTM unit has five components: an input gate, a hidden state, a memory cell, a forget gate, and output gate. The memory cell unit combines the previous time step memory cell unit which is modulated by the forget and input modulation gates together with the previous hidden state, modulated by the input gate. These components enable an LSTM to learn very complex long-term and temporal dynamics that a vanilla RNN is not capable of. Additional depth in capturing the complexity of a tie series can be added by stacking LSTM on top of each other. 

Formally, at each time step a new memory cell $c_t$ is created taking current input $z_t$ and previous hidden state $h_{t-1}$ and it is then combined with forget gate that controls an amount of information kept in the hidden state as
\begin{eqnarray*}
h_{t} &=& \sigma\left(\underbrace{W_h^{(o)} h_{t-1} + W_z^{(o)} =z_t + b_0^{(o)}}_{\text{output gate}}\right) \circ \tanh (c_t) \\
c_{t} &=& \sigma\left(\underbrace{W_h^{(g)} h_{t-1} + W_z^{(g)} z_t + b_0^{(g)}}_{\text{forget gate}}\right) \circ c_{t-1} + 
\sigma\left(\underbrace{W_h^{(i)} h_{t-1} + W_z^{(i)} + b_0^{(i)}}_{\text{input gate}} \right) \circ \tanh(k_t).
\label{eq: RRnet}
\end{eqnarray*}
The term $\sigma(\cdot)\circ c_{t-1}$ introduces the long-range dependence, $k_t$ is new information flow to the current cell. The forget gate and input gate states control weights of past memory and new information. In the Figure \ref{fig: RNnet}, $c_t$ is the memory pass through multiple hidden states in the recurrent network.

\subsection{Loss Function}
\label{sec: loss}

Since we aim to estimate the cumulative distribution function that is a non-decreasing function bounded on $[0,1]$, we need to design an objective function that minimizes differences between targets and estimated distribution as well as imposes non-decreasing property of the output. Since the problem is essentially a more complex classification problem closely related to logistic regression, we use a binary cross-entropy loss function. Moreover, to order the predicted probabilities, we introduce a penalty to the multiple output classification problem.

The loss function is then composed of two parts: traditional binary cross-entropy and a penalty imposing monotonicity of predicted output:
\begin{eqnarray}
\label{eq: loss-bince}
   \nonumber \mathcal{L} &=& \frac{1}{T}\sum_t^{T} \underbrace{\frac{1}{p} \sum_j^p\left(\mathbb{I}_{\{y_{t+h}\le q^{\alpha_j}\}}\log\left\{\widehat{\mathfrak{g}}_{W,b,j}(z_t)\right\} + \left(1-\mathbb{I}_{\{y_{t+h}\le q^{\alpha_j}\}}\right)\log\left\{1-\widehat{\mathfrak{g}}_{W,b,j}(z_t)\right\} \right)}_\text{\text{binary cross-entropy}} \\
    &&+~ \underbrace{\lambda_{m} \sum_t^{T} \left(\sum_{j=1}^{p-1}\left(\widehat{\mathfrak{g}}_{W,b,j}(z_t) - \widehat{\mathfrak{g}}_{W,b,j+1}(z_t)\right)\right)_+}_\text{\text{monotonicity penalty}}
\end{eqnarray}
where $(u)_+$ is a rectified linear units function, ReLU, $(u)_+ = \max\{u,0\}$, which passes through only positive differences between two neighboring values, $j$ and $j+1$, of CDF, those violating the monotonicity condition, and $\mathbb{I}_{\{.\}}$ is an indicator function. This violation is controlled by the penalty parameter $\lambda_m$. Note that in addition to its simplicity, ReLU is used for convenience reasons allowing for general use.\footnote{This choice allows to use GPU and hence opens computational capacities for more complex problems. The use of own or not optimized functions for GPU is not desired and $(u)_+$ is common to libraries working with GPUs.}


\subsection{Networks Design and Estimation Steps}
\label{sec: estimation_steps}

Due to the high dimensionality and non-linearity of the problem, estimation of a deep neural network is a complex task. Estimation requires optimal selection of parameters to provide good performance and avoid potential risks such as over-fitting or convergence problems. Moreover, each problem and data require specific careful choices to minimize the risks. Here, we provide a detailed summary of the model architectures and their estimations.

\subsubsection{Learning, Regularization and Hyper-parameters}
Selection of hyper-parameters together with regularization methods play crucial role in reduction of risk from estimation. In particular, we use ReLU activation function to introduce non-linearity to our problem, and help the optimization algorithm converge faster. For the learning process, we use adaptive gradient algorithm, Adam \citep{kingma2014adam} and its modification AdamW \citep{loshchilov2019adamw} that allows for regularization by decoupling the weight decay from the gradient-based update. The regularization is close to $L_2$-regularization with improved results.\footnote{We keep decay of momentum parameters $\beta_1, \beta_2$ constant and at default values throughout all estimations, $\beta_1 = 0.9$, $\beta_2=0.999$.}

We use hyper-optimization algorithms based on random search over a grid/cube of parameter ranges, which are specific to a given experiment. For some experiments one might search over the full grid of parameters testing all possible combinations, which can be costly. Using hyper-optimization\footnote{We using Julia package HyperOpt.jl (\url{https://github.com/baggepinnen/Hyperopt.jl}).} we select the learning rate of the optimizer, $\eta$, the weight decay parameter of AdamW, $\lambda_W$, and a Dropout parameter regularizing models \citep{srivastava2014dropout} that is an efficient way of performing model averaging with neural networks. Specifically, Dropout parameter turns-off a fraction $\phi \in (0.0,1.0)$ of nodes in the layer of network at which it is applied. In an in-sample training part, the model is given number of epochs to learn on the data. The number of epochs depend on the size of the data and the batch size used in the estimation. However, we also use early stopping technique that helps regularization to prevent from over-fitting. The early-stopping, or patience, criterion is the minimum number of epochs provided to model to learn.

\subsubsection{Code Implementation}

We have estimated our models on 48 core Intel\textregistered~ Xeon\textregistered~/i7 Gold 6126 CPU@ 2.60GHz, 128 GB of memory, and GeForce 3090 GPU. We implement the models using \texttt{Flux.jl} \citep{Flux2018} package in \textsf{JULIA} 1.6.0. language.

\subsubsection{Data Preparation and Information set}

To predict the distribution function of a time series ${y_t}$ with observations $y_1, \dots, y_T$, we split our time series into several parts. The first partitioning creates, as known in time-series literature, in-sample $[1:t_0]$ and out-of-sample $[t_0+1:T]$ subsamples. Equivalently, in machine learning jargon, train and test sets. Test subsample is never available to the learning algorithm while training the model. We further divide the train subsample into training and validation sets, which are used to cross-validation of our model and model's parameters selection. The model selection is based on the value of the loss function on the validation subsample(s), mainly the binary cross-entropy loss.

One of the crucial parts in estimation of distributional neural networks is the information set. The information set $\mathcal{I}_{t_0}$ is based on the past observation available at time $t_0$. This is the maximal time-span providing historical information, in our case, up to the last observation of the validation subsample. The importance here lays in finding the empirical quantiles, $q^{\alpha}$, corresponding to the set of probabilities $\{\alpha_{1},\ldots,\alpha_{p}\}$, which are used to build the sequence of target values. Given the information about $\{y_{t}\}_{t=1}^{t_0}$ and the empirical quantiles, we are able to model the distribution conditional on the information set up to time $t_0$. Given the information set, we face the problem with non-variation or updating the conditional empirical quantiles for the future distributions. Although, we do not assume a shape of the distribution, we assume, to some extent, small level of shift in the distribution.
Further, choice of empirical quantiles and probability levels faces the same problem as quantile regression when it comes to small samples of data. 

\section{Empirical Application: Macroeconomic Fan Charts in Era of Big Data}
\label{sec: empirics-macro}

Macroeconomic fan charts are popular tool for communicating uncertainty surrounding economic forecasts. Recognizing the need of communicating uncertainty to public, the Bank of England started to publish fan charts in 1996 and quicky became a leader in communicating uncertainty. Yet the art and science of such important tool for policy making remains on shoulders of the methods chosen.

Here we aim to construct a data-driven macroeconomic fan charts from a best approximating model learned from hundreds of variables with deep learning. This is in sharp contrast to the literature providing uncertainty of macroeconomic variables using so-called prediction fan charts \citep{britton1998inflation,stock2017twenty} using few variables with parametrized and structured model that requires number of assumptions.

\subsection{Data} 

To construct such tool for measurement of uncertainty, we use a high-dimensional dataset of \cite{mccracken2020fred} that has been extensively used in the macroeconomic literature \citep{coulombe2020machine} and is available at the Federal Reserve of St-Louis's web site. From several alternatives, we opt for the harder to predict quarterly data FRED-QD. Our dataset contains 216 quarterly US macroeconomic and financial indicators observed from 1961Q1 until 2019Q4. Since number of variables are non-stationary, we follow the transformation codes used by \cite{mccracken2020fred}. Using this dataset we construct a data-rich fan charts for real GDP growth (GDPC1), inflation (CPIAUCSL), unemployment rate (UNRATE) that will to the best of our knowledge be the first of its kind to reflect high-dimensional information of 216 relevant variables. 
To contrast the data-driven fan charts, we use an state-of-the-art macroeconomic model based on Bayesian Vector Autoregression that contains data-driven factors to incorporate the big data information \citep{mccracken2020fred}. 

\subsection{Deep-learning Based Fan Charts} 
In order to obtain an $h$-step ahead forecasts that will form a fan chart, we consider direct prediction scheme. Exploring the data structures, we form $h$ distributional networks
\begin{equation}
    \widehat{\mathfrak{g}}_{W,b}^{(1)}(z_t),\ldots,\widehat{\mathfrak{g}}_{W,b}^{(h)}(z_t),
\end{equation}
where entire (continuous) $h$-step-ahead conditional distribution $\widehat{\mathfrak{g}}_{W,b}^{(h)}(z_t)$ is obtained by interpolation of cumulative distribution function preserving the monotonicity of the outcome. Here we apply Fritsch-Carlson monotonic cubic interpolation \citep{fritsch1980interpolation}, for details see Appendix \ref{app:interpolation} and use the predicted cumulative distribution function $\widehat{F}_{t+h}(\alpha|\mathcal{I}_t)$ to form $k-size$ prediction intervals for a fan chart as
 \begin{equation}
PI_{t+h}^{k} = \left[\widehat{F}^{-1}_{t+h}(\alpha_l|\mathcal{I}_t), \widehat{F}^{-1}_{t+h}(\alpha_u|\mathcal{I}_t)\right],
\label{eq:predinterval}
\end{equation}
such that $1-k = \alpha_u - \alpha_l$ is size of the interval. 

To show how useful our approach is, we contrast the predictions obtained from distributional network with the Bayesian vector auto-regression estimated on the factors extracted from data as in \cite{mccracken2020fred}. This benchmark is a state-of-the-art approach in macroeconomics and at the same time uses information from the whole dataset, hence the forecasts are comparable. To obtain fan chart (prediction intervals), we use the best performing recursive (iterative) scheme $\widehat{y}_{t+h} = f(y_{t+h-1}| \mathcal{I}_t)$ where the prediction intervals are based on the distribution of residuals. Formally, when we assume normal distribution we obtain $h$-step ahead $\alpha$ prediction interval as $\left[\widehat{y}_{t+h} - \phi(1-\alpha/2) \widehat{\sigma}_h, ~ \widehat{y}_{t+h} + \phi(1-\alpha/2) \widehat{\sigma}_h\right]$, where $\phi(1-\alpha/2)$ is corresponding quantile of the std. normal distribution, and, for example, for a naive forecast we have $\hat{\sigma}_h = \hat{\sigma}\sqrt{h}$ and $\hat{\sigma}_h$ is the residual standard deviation.\footnote{Alternatively, one can obtain prediction intervals or fan charts using Bootstrap methods, \citet{britton1998inflation}.}

We evaluate the $h$-step-ahead forecasts with quantile loss function \citep{clements2008quantile} 
\begin{equation}
	L_{\alpha, m}^h = E (\alpha - \mathbb{I}\{e_{t+h,m} < 0\}) e_{t+h,m}),
    \label{eq:tickloss}
\end{equation}
for a model $m$, horizon $h$, and ${\alpha}$-quantile where $e_{t+h,m} = y_{t+h} - \widehat{F}^{-1}_{t+h,m}(\alpha|\mathcal{I}_t)$ is the difference between the original time series and $\alpha$-quantile forecast given the information set, $\mathcal{I}_t$. To compare the predictive accuracy of the models, we use the Diebold-Mariano test \citep{diebold1995comparing} with Newey-West variance for $h>1$ cases and test the null hypothesis $H_0: L_{\alpha,m_1} > L_{\alpha,m_2}$ against the alternative that $m_2$ is less accurate than $m_1$. 

\subsection{Setup}

Working with quarterly data, we compute $h=1,\dots,6$ horizon forecasts each quarter of the out-of-sample period starting at 2012:Q3 and ending with 2019:Q4. Conditional distribution is approximated with $j=1,\ldots,19$ empirical $\alpha_j = (0.01, ..., 0.99)$ probability levels. The learning explores 36 combinations of hyper-parameters to find the best approximating model for each $h$-step ahead forecast separately. The hyper-parameters space is optimized once on the training part of the data, and the training procedure performs growing-window forward-validation scheme on the training data using 3-folds. We split the training data while training each fold of validation on train and test parts by ratio $0.93$. We present predictions for deep recurrent neural networks with two hidden LSTM layers of different numbers of neurons chosen in the hyper-optimization. Table~\ref{Tab:pars-macro} in Appendix summarizes all parameters and details used in the estimation. To compare the deep-learning based fan charts, we perform the standard estimation procedure for the Bayesian Vector Autoregression (BVAR) model with factor components as in \citet{mccracken2020fred}. We use the information criteria\footnote{AIC, BIC} and choose the model with four lags. Further, we find the prediction intervals for GDP growth (GDP), inflation (CPI), and the unemployment rate (UNE). The data for both procedures are transformed according to \citet{mccracken2020fred} codes and standardized to normal with zero mean and standard deviation one.

\subsection{Discussion} 

\begin{figure}[!ht]
    \centering
    \includegraphics[width=0.99\textwidth,trim=0cm 0cm 0cm 0cm, clip=true]{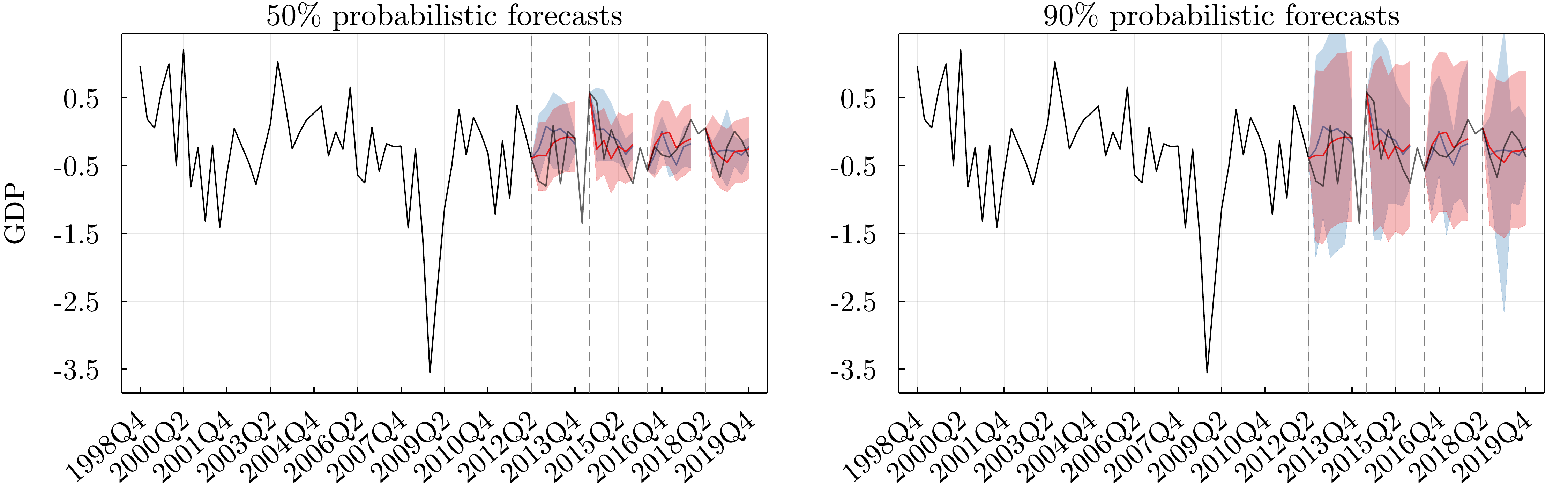}
    \includegraphics[width=0.99\textwidth,trim=0cm 0cm 0cm 0cm, clip=true]{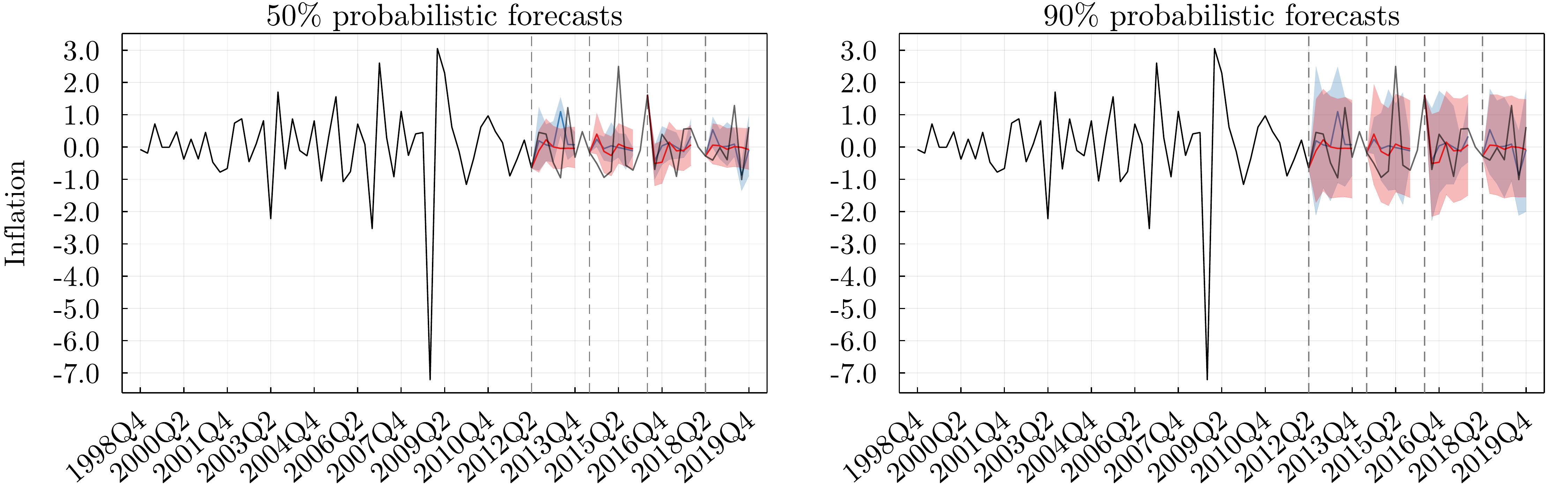}
    \includegraphics[width=0.99\textwidth,trim=0cm 0cm 0cm 0cm, clip=true]{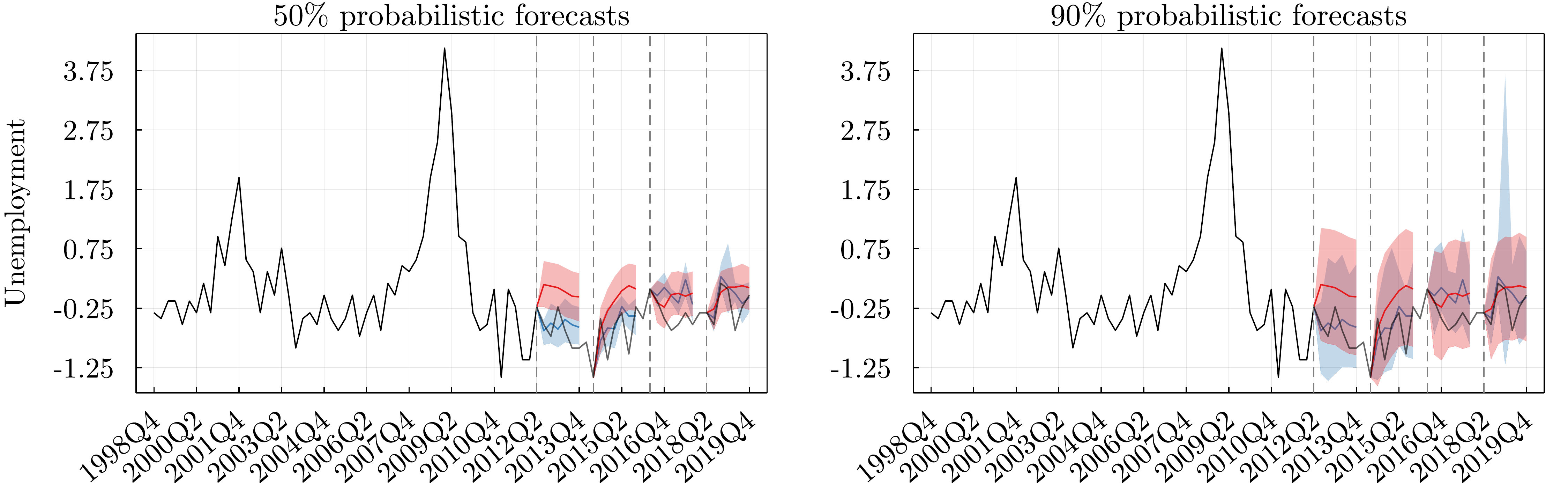}
    \caption{Deep-learning based (blue) and BVAR (red) fan charts }
    \begin{minipage}{\textwidth} 
            \footnotesize
            6-step-ahead quarterly forecasts of GDP growth (top), Inflation (middle), and Unemployment rate (bottom) with 50\% (left column) and 90\% (right column) fan charts obtained by distributional network (blue) using 216 quarterly US macroeconomic and financial indicators from the FRED-QD database, and a factor three+four-variable BVAR (red). Forecasts are made at the end of the 2012:Q2, 2014:Q2, 2016:Q2 and 2018:Q2 depicted by dashed vertical lines. Train data are plotted by black solid line and test data by grey solid line.
        \end{minipage} 
    \label{fig:macro-fancharts3}
\end{figure}

We start the discussion by presenting the qualitative results of gdp growth, inflation and unemployment predictions in form of fan charts. Figure~\ref{fig:macro-fancharts3} compares median as well as 50\% and 90\% prediction intervals made at four different periods by both recurrent distributional neural network and bayesian vector auto-regression approaches and highlights the benefits of deep learning-based predictions.

The prediction intervals from distributional neural network are asymmetric, and in contrast to traditional time series represented by BVAR are not very smooth over forecast horizon. With growing uncertainty when looking ahead into future the deep learning still learns some structure from data and probability intervals resemble less the form of ``fans''. The intervals are narrower in comparison to BVAR in most of the cases, especially when looking at 50\% intervals. In case of unemployment rate, the BVAR model is less capable to reduce the uncertainty about the future observation most probably because of strong spikes in previous years. In contrast, our DistrNN with LSTM units captures the uncertainty precisely.

While Figure~\ref{fig:macro-fancharts3} is illustrative, it only shows few periods and to support the gains of deep learning approach we further quantify the prediction differences for whole out-of-sample period. Tables~\ref{tab:quantilelossesmacro} and~\ref{tab:dieboldmarianomacro} present the quantitative comparison of predictions from both models. We compare the forecasts at $h=1,\dots,6$ horizons using tick loss (Eq. \ref{eq:tickloss}) for selected $\alpha=\{0.1, 0.25, 0.5, 0.75, 0.9\}$ probability levels.

In Table~\ref{tab:quantilelossesmacro} we report quantile loss of both Distributional Recurrent Neural Network (DistrNN) and Bayesian VAR (BVAR) forecasts of GDP growth (GDP), inflation, and unemployment rate. Deep-learning based DNN approach provides forecasts with lower error (in blue) at most considered probability levels and horizons and it brings larger improvement at shorter horizons. Notable gains are at 90\% where DistrNN dominates BVAR greatly. With exception of inflation, DNN also improves losses for median and 25\% level forecasts.

\begin{table}[!ht]
  \footnotesize
  \centering
  \caption{Quantile loss of DistrNN and BVAR } 
    \begin{tabular}{rllllllllllllll}
    \toprule
         & \multicolumn{2}{c}{\textbf{0.1}} &   & \multicolumn{2}{c}{\textbf{0.25}} &   & \multicolumn{2}{c}{\textbf{0.5}} &   & \multicolumn{2}{c}{\textbf{0.75}} &   & \multicolumn{2}{c}{\textbf{0.9}} \\
\cmidrule{2-3}\cmidrule{5-6}\cmidrule{8-9}\cmidrule{11-12}\cmidrule{14-15}    
   & \multicolumn{1}{c}{DistrNN} & \multicolumn{1}{c}{BVAR} &   & \multicolumn{1}{c}{DistrNN} & \multicolumn{1}{c}{BVAR} &   & \multicolumn{1}{c}{DistrNN} & \multicolumn{1}{c}{BVAR} &   & \multicolumn{1}{c}{DistrNN} & \multicolumn{1}{c}{BVAR} &   & \multicolumn{1}{c}{DistrNN} & \multicolumn{1}{c}{BVAR} \\
\cmidrule{2-3}\cmidrule{5-6}\cmidrule{8-9}\cmidrule{11-12}\cmidrule{14-15}   
GDP    &   &   &   &   &   &   &   &   &   &   &   &   &   &  \\
       $h=1$ & \textcolor{myblue}{\textbf{1.985}} & 2.146 &   & \textcolor{myblue}{\textbf{2.829}} & 3.645 &   & \textcolor{myblue}{\textbf{3.751}} & 4.280 &   & 3.118 & 3.060 &   & \textcolor{myblue}{\textbf{1.917}} & 2.381 \\
       $h=2$ & \textcolor{myblue}{\textbf{2.253}} & 2.305 &   & \textcolor{myblue}{\textbf{3.338}} & 3.478 &   & 3.957 & 3.671 &   & \textcolor{myblue}{\textbf{2.800}} & 3.026 &   & \textcolor{myblue}{\textbf{1.826}} & 2.665 \\
       $h=3$ & 2.317 & 2.193 &   & \textcolor{myblue}{\textbf{3.444}} & 3.510 &   & \textcolor{myblue}{\textbf{4.249}} & 4.307 &   & 3.524 & 3.203 &   & \textcolor{myblue}{\textbf{2.287}} & 2.363 \\
       $h=4$ & \textcolor{myblue}{\textbf{1.977}} & 2.130 &   & \textcolor{myblue}{\textbf{3.437}} & 3.626 &   & \textcolor{myblue}{\textbf{4.013}} & 4.513 &   & \textcolor{myblue}{\textbf{2.931}} & 3.349 &   & \textcolor{myblue}{\textbf{1.710}} & 2.679 \\
       $h=5$ & \textcolor{myblue}{\textbf{1.843}} & 2.199 &   & \textcolor{myblue}{\textbf{3.141}} & 3.193 &   & \textcolor{myblue}{\textbf{3.850}} & 4.222 &   & \textcolor{myblue}{\textbf{2.564}} & 3.188 &   & \textcolor{myblue}{\textbf{1.549}} & 2.546 \\
       $h=6$ & 2.176 & 2.159 &   & 3.298 & 3.036 &   & \textcolor{myblue}{\textbf{3.776}} & 4.066 &   & \textcolor{myblue}{\textbf{2.682}} & 3.061 &   & \textcolor{myblue}{\textbf{1.909}} & 2.374 \\
\cmidrule{2-3}\cmidrule{5-6}\cmidrule{8-9}\cmidrule{11-12}\cmidrule{14-15}    
\multicolumn{2}{l}{Inflation}      &   &   &   &   &   &   &   &   &   &   &   &   &  \\
       $h=1$ & \textcolor{myblue}{\textbf{4.017}} & 4.051 &   & \textcolor{myblue}{\textbf{6.294}} & 6.884 &   & \textcolor{myblue}{\textbf{7.711}} & 9.509 &   & \textcolor{myblue}{\textbf{7.373}} & 8.552 &   & \textcolor{myblue}{\textbf{4.472}} & 7.925 \\
       $h=2$ & 3.408 & 3.102 &   & 6.339 & 5.864 &   & 8.554 & 8.411 &   & \textcolor{myblue}{\textbf{7.325}} & 8.611 &   & \textcolor{myblue}{\textbf{3.959}} & 8.650 \\
       $h=3$ & \textcolor{myblue}{\textbf{2.509}} & 3.171 &   & 5.319 & 5.295 &   & 8.394 & 7.937 &   & 8.291 & 8.252 &   & \textcolor{myblue}{\textbf{5.574}} & 8.453 \\
       $h=4$ & \textcolor{myblue}{\textbf{2.899}} & 2.955 &   & 5.889 & 5.596 &   & 8.481 & 8.399 &   & \textcolor{myblue}{\textbf{7.485}} & 8.169 &   & \textcolor{myblue}{\textbf{5.023}} & 8.210 \\
       $h=5$ & 3.454 & 3.037 &   & 5.973 & 5.261 &   & 8.004 & 7.944 &   & \textcolor{myblue}{\textbf{7.138}} & 8.102 &   & \textcolor{myblue}{\textbf{4.941}} & 8.101 \\
       $h=6$ & 4.004 & 3.055 &   & 6.747 & 5.215 &   & 8.198 & 7.850 &   & \textcolor{myblue}{\textbf{6.941}} & 8.006 &   & \textcolor{myblue}{\textbf{4.645}} & 8.178 \\
\cmidrule{2-3}\cmidrule{5-6}\cmidrule{8-9}\cmidrule{11-12}\cmidrule{14-15}    
\multicolumn{2}{l}{Unemployment}      &   &   &   &   &   &   &   &   &   &   &   &   &  \\
       $h=1$ & \textcolor{myblue}{\textbf{1.950}} & 2.182 &   & \textcolor{myblue}{\textbf{3.060}} & 3.718 &   & \textcolor{myblue}{\textbf{3.586}} & 4.799 &   & \textcolor{myblue}{\textbf{2.928}} & 3.495 &   & \textcolor{myblue}{\textbf{1.639}} & 2.634 \\
       $h=2$ & \textcolor{myblue}{\textbf{1.860}} & 2.359 &   & \textcolor{myblue}{\textbf{3.126}} & 4.022 &   & \textcolor{myblue}{\textbf{3.703}} & 4.814 &   & \textcolor{myblue}{\textbf{2.995}} & 3.155 &   & \textcolor{myblue}{\textbf{1.794}} & 2.168 \\
       $h=3$ & \textcolor{myblue}{\textbf{1.902}} & 1.956 &   & \textcolor{myblue}{\textbf{3.407}} & 3.753 &   & \textcolor{myblue}{\textbf{4.004}} & 5.128 &   & 3.579 & 3.563 &   & \textcolor{myblue}{\textbf{2.486}} & 2.614 \\
       $h=4$ & 2.605 & 1.841 &   & 3.712 & 3.666 &   & \textcolor{myblue}{\textbf{3.884}} & 5.600 &   & \textcolor{myblue}{\textbf{3.252}} & 3.769 &   & \textcolor{myblue}{\textbf{2.193}} & 2.735 \\
       $h=5$ & 2.187 & 1.910 &   & \textcolor{myblue}{\textbf{3.851}} & 3.851 &   & \textcolor{myblue}{\textbf{4.383}} & 5.777 &   & \textcolor{myblue}{\textbf{3.263}} & 3.851 &   & \textcolor{myblue}{\textbf{1.816}} & 2.621 \\
       $h=6$ & 2.113 & 1.916 &   & \textcolor{myblue}{\textbf{3.253}} & 3.751 &   & \textcolor{myblue}{\textbf{3.731}} & 5.776 &   & \textcolor{myblue}{\textbf{2.929}} & 3.728 &   & \textcolor{myblue}{\textbf{1.756}} & 2.518 \\
    \bottomrule
    \end{tabular}%
    \\[0.5em]
    \begin{minipage}{\textwidth}
        \footnotesize
        \textbf{Note:} Quantiles losses of Distributional Recurrent Neural Network (DistrNN) and Bayesian VAR (BVAR), for variables GDP growth (GDP), Inflation, and Unemployment rate. The out-of-sample forecasts for 25 quarters are made at $\alpha$-levels $\{0.1, 0.25, 0.5, 0.75, 0.9\}$, $horizons h=1,\dots,6$,  starting at Q3/2013 and ending at Q4/2019. Cases with DNN forecast being smaller in comparison to BVAR are in blue.
    \end{minipage}%
    \label{tab:quantilelossesmacro}%
\end{table}%

While deep-learning provides us with better forecasts in most of the cases, number of these cases is also significantly different from a traditional BVAR approach. Table~\ref{tab:dieboldmarianomacro} reports the results supporting the relative performance of the two methods. Deep learning delivers significantly better prediction in comparison to BVAR in 12 cases while BVAR never outperforms the deep-learning approach significantly with exception of 5-step-ahead forecast of inflation at 25\% quantile level. We note that the results depend on 25 out-of-sample observations, which is a size that might be limiting for the test.

\begin{table}[!ht]
	\footnotesize
  \centering
  \caption{Relative out-of-sample performace of DistrNN and BVAR} 
    \begin{tabular}{rcllllllllll}
    \toprule
      &   &   & \textbf{0.1} &   & \textbf{0.25} &   & \textbf{0.5} &   & \textbf{0.75} &   & \textbf{0.9} \\
\cmidrule{4-4}\cmidrule{6-6}\cmidrule{8-8}\cmidrule{10-10}\cmidrule{12-12}    GDP &   &   &   &   &   &   &   &   &   &   &  \\
      & $h=1$ &   & -0.4234 &   & \textcolor{myblue}{\textbf{-2.0609 *}} &   & -0.8421 &   & 0.1479 &   & -1.0619 \\
      & $h=2$ &   & -0.1094 &   & -0.4939 &   & 0.5027 &   & -0.2906 &   & -0.9199 \\
      & $h=3$ &   & 0.1516 &   & -0.0821 &   & -0.0757 &   & 0.5351 &   & -0.0948 \\
      & $h=4$ &   & -0.8808 &   & -1.6944 &   & -0.6268 &   & -0.9966 &   & -1.1122 \\
      & $h=5$ &   & -1.4336 &   & -0.4723 &   & -0.4374 &   & \textcolor{myblue}{\textbf{-2.5705 **}} &   & -1.3932 \\
      & $h=6$ &   & 0.0508 &   & 0.8273 &   & -0.3862 &   & -0.6077 &   & -0.5314 \\
\cmidrule{4-4}\cmidrule{6-6}\cmidrule{8-8}\cmidrule{10-10}\cmidrule{12-12}    \multicolumn{2}{l}{Inflation}   &   &   &   &   &   &   &   &   &   &  \\
      & $h=1$ &   & -0.0369 &   & -0.5153 &   & -1.6002 &   & -0.8522 &   & \textcolor{myblue}{\textbf{-1.718 *}} \\
      & $h=2$ &   & 0.7832 &   & 0.5475 &   & 0.1647 &   & -1.1302 &   & \textcolor{myblue}{\textbf{-3.1631 ***}} \\
      & $h=3$ &   & \textcolor{myblue}{\textbf{-4.7111 ***}} &   & 0.0436 &   & 1.0352 &   & 0.0345 &   & \textcolor{myblue}{\textbf{-2.1701 **}} \\
      & $h=4$ &   & -0.3859 &   & 0.3249 &   & 0.151 &   & -0.9863 &   & \textcolor{myblue}{\textbf{-1.7402 *}} \\
      & $h=5$ &   & 0.5635 &   & \textcolor{myred}{\textbf{3.9307 ***}} &   & 0.101 &   & \textcolor{myblue}{\textbf{-5.553 ***}} &   & \textcolor{myblue}{\textbf{-14.7935 ***}} \\
      & $h=6$ &   & 0.8104 &   & 1.5175 &   & 0.2902 &   & -1.0714 &   & \textcolor{myblue}{\textbf{-2.1468 **}} \\
\cmidrule{4-4}\cmidrule{6-6}\cmidrule{8-8}\cmidrule{10-10}\cmidrule{12-12}    \multicolumn{2}{l}{Unemployment}   &   &   &   &   &   &   &   &   &   &  \\
      & $h=1$ &   & -0.3937 &   & -0.8682 &   & \textcolor{myblue}{\textbf{-1.7259 *}} &   & -1.0862 &   & -1.2746 \\
      & $h=2$ &   & -0.9097 &   & -1.3733 &   & -1.5014 &   & -0.2501 &   & -0.4865 \\
      & $h=3$ &   & -0.1361 &   & -0.4044 &   & -1.4017 &   & 0.0155 &   & -0.1076 \\
      & $h=4$ &   & 0.8105 &   & 0.0321 &   & -1.3888 &   & -0.4531 &   & -0.5153 \\
      & $h=5$ &   & 1.479 &   & -0.0008 &   & -1.1089 &   & -0.7159 &   & -1.1653 \\
      & $h=6$ &   & 1.1521 &   & -0.6429 &   & \textcolor{myblue}{\textbf{-2.0862 **}} &   & -1.7029 &   & -1.6384 \\
    \bottomrule
    \end{tabular}%
    \\[0.5em]
    \begin{minipage}{\textwidth}
        \footnotesize
        \textbf{Note:} The values are Diebold-Mariano test statistics, with the null hypothesis $H_0: L_{\alpha,DistrNN} > L_{\alpha,BVAR}$ against the alternative that Bayesian VAR  is less accurate than the Distributional Recurrent Neural Network. Stars indicate statistical significance that ***, **, * correspond to 1\%, 5\%, 10\% levels, accordingly. Negative sign of the DM statistics states that DistrNN has better OOS performance than BVAR, positive sign states opposite. We report the out-of-sample forecasts for 25 quarters for three variables GDP growth (GDP), Inflation, and Unemployment rate, at $\alpha$-levels $\{0.1, 0.25, 0.5, 0.75, 0.9\}$, horizons $h=1,\dots,6$, starting at Q3/2013 and ending at Q4/2019. 
    \end{minipage}%
  \label{tab:dieboldmarianomacro}%
\end{table}%

\clearpage
\section{Empirical Application: Conditional Distributions of Asset Returns}
\label{sec: results-assets}

Stock returns data are notoriously known to contain heavy tails and low signal-to-noise ratio \citep{fama1965portfolio,israel2020can}. Despite the large literature uncovering these empirical properties, only few studies attempt to forecast the distribution of returns.\footnote{Literature focusing on Value-at-Risk forecasting has a special interest in a chosen quantile of the return distribution, mostly left tail \citep{engle2004caviar}} Among the few, \cite{barunik2019simple} parametrize a simple ordered logit to deliver the distribution forecasts. 

Here we aim to build a machine-learning based alternative that is capable of exploring large number of informative variables. We compare the forecasts to the benchmark \cite{barunik2019simple} (henceforth AB) model to see how well the machine learning approach approximates the parametrized model, but we mainly focus on using more variables that classical models can not explore due to its in-feasibility connected with large parameter space. 

This application hence serves as a good complement to the previous one where we used machine learning for multiple variable forecast using big data. In contrast, liquid stock returns are known to be hardly predictable, and hence even small improvement is valuable.

\subsection{Data and Estimation}

Our dataset includes 29 most liquid U.S. stocks\footnote{Assets selected in the sample: AAPL, AMZN, BAC, C, CMCSA, CSCO, CVX, DIS, GE, HD, IBM, INTC, JNJ, JPM, KO, MCD, MRK, MSFT, ORCL, PEP, PFE, PG, QCOM, SLB, T, VZ, WFC, WMT, XOM.} of S\&P500. The main reason for this particular choice is comparability of the results with \citet{barunik2019simple}. The daily data covers the period from July 1, 2005 to August 31, 2018. We preprocessed the data to eliminate possible problems with liquidity or biases caused by weekend or bank holidays. The final sample period contains 3261 observations. 

We start building the models using the same predictors as in \citet{barunik2019simple} to make a direct comparison of the model forecasts. Specifically, they use $\text{Ind}_t=\mathbb{I}_{\{r_{t} \leq q^{\alpha_j}\}}$ and $\text{LogVol}_t=\ln(1 + |r_{t}|)$ as a proxy to a volatility measure. We will refer to this first choice as the \textit{AB predictors}. Next, we prepare five realized measures from one-minute intra-day high frequency data obtained from TickData.\footnote{\url{www.tickdata.com}} The realized measures for each of 29 asset returns are realized volatility, skewness, kurtosis, and positive and negative semi-variances labelled as $\text{RVol}_t$, $\text{RSkew}_t$, $\text{RKurt}_t$, $\text{RSemiPos}_t$, and $\text{RSemiNeg}_t$. These are informative about returns distribution and should help the forecast. We will refer to this set of as \textit{RM predictor}
\begin{table}[ht]
    \centering
    \begin{tabular}{ l l l}
    \toprule
    AB predictors & RM predictors & AB+RM predictors \\ 
    \midrule
    $\text{Ind}_t$ &  -  & $\text{Ind}_t$  \\
    $\text{LogVol}_t$  &  -  & $\text{LogVol}_t$  \\
    -         &  $\text{RVol}_t$  & $\text{RVol}_t$  \\
    -         &  $\text{RSkew}_t$    & $\text{RSkew}_t$  \\
    -         &  $\text{RKurt}_t$    & $\text{RKurt}_t$  \\
    -         &  $\text{RSemiPos}_t$  & $\text{RSemiPos}_t$  \\
    -         &  $\text{RSemiNeg}_t$  & $\text{RSemiNeg}_t$  \\
    \bottomrule
    \end{tabular}
    \caption{Sets of predictors used in the three models}
\begin{minipage}{\textwidth} 
\centering
            \footnotesize
            Note: The indicator $\text{Ind}_t$ contains $J$ columns of dummy variables.
        \end{minipage} 
    \label{logitdata}
\end{table}

In the third model, we combine both sets of predictors to estimate the conditional distribution of return since they are both informative. Since the realized measures contain information about higher moments of returns distribution these might improve predictions of the conditional distributions of returns. At the same time,  inclusion of those predictors into the original (benchmark) ordered logit model of \cite{barunik2019simple} would result in over-parametrized model that is not feasible. This is an important note, since our approach provides flexible and more general way of predicting distributions in data-rich environment, and at the same time explores possible non-linearity in data. Table \ref{logitdata} summarizes the predictors used in the three models.

Prior to the estimation, we normalize the input data to an appropriate range which eases the job for the algorithm to find better optimum. This is a standard procedure in learning process since the optimization operates on closer ranges while learning in the network structure. Further, we split the data into train and test parts with ratio 0.9 and 0.1, specifically to 2934 and 327 observations respectively. First, we search for the best hyper-parameters set on the training window, at which we perform four-fold rolling-window forward-validation scheme. The model is trained and validated during hyper-parameter search on each split composed from 90\% and 10\% partitions - training and validation. Using rolling window of size 2934, we predict one step-ahead out-of-sample forecasts, H = 1. The window size equals to the size of the training sample, $t_0=2934$. On the first rolling window, training part, we search the grid the best parameters set using Random and Latin hypercube search algorithms. Table~\ref{Tab:pars-assets} in appendix \ref{app:tables} details the ranges of parameters for the learning rate, $\eta$, dropout parameter, $\phi$, and the weight decay penalizing parameter, and $\lambda_W$, on which the hyper-optimization algorithm is searching for the best model hyper-parameters in the space of 50 combinations. We employ the ensemble method for prediction, thus, for each rolling window step, the best model is trained three times given the best model hyper-parameters. Three forecasts are obtained and an average distribution forecast is made for all $t$ in out-of-sample, $t\in[2935:3261]$. In addition, we use additional regularization technique of early stopping and Table~\ref{Tab:pars-assets} in appendix \ref{app:tables} also provides number of epochs allowed to train. This is number of epochs the model is patient about algorithm and waiting on improvement. Finally, we take the model with best validation loss and use it for prediction of out-of-sample distributions.
We also study an effect of complexity specifying the size of neural networks. We set the number of nodes from a shallow to deeper DistrNN to $[128]$, $[128,64]$, and $[128,64,32]$. The network's final layer outputs size is $p=10$, which values correspond to probability forecasts approximating the conditional distribution of excess returns given the information set $\mathcal{I}_t$.

\subsection{Statistical Evaluation Measures}

We evaluate our probabilistic forecasts using several measures. First, we evaluate the precision of forecasts using the mean square prediction error calculated as
\begin{equation}
\label{eq: loss-mse}
    MPSE = \frac{1}{T-t_0} \sum_{t=t_0+1}^{T} \frac{1}{p} \sum_{j=1}^p\left(\mathbb{I}_{\{y_{t+h}\le q^{\alpha_j}\}} - \widehat{\mathfrak{g}}_{W,b,j}(z_t)\right)^2,
\end{equation}
where the out-of-sample predicted outputs $\widehat{\mathfrak{g}}_{W,b,j}(z_t)$ is a matrix keeping dimension of time $[t_0+1,\dots,T]$ and conditional probability levels $\{1,\dots,p\}$.

To evaluate the compatibility of a cumulative distribution functions with an individual time series observations we use the continuous ranked probability score (CRPS, \citet{matheson1976scoring}, \citet{hersbach2000decomposition}):
\begin{equation}
\label{eq: crps}
CRPS_t = -\int_{-\infty}^{\infty} \left( \widehat{\mathfrak{g}}_{W,b}(z_t) - \mathbb{I}_{\{y_{t+h}\le y\}} \right)^2 d y,
\end{equation}
where the conditional CDF $\widehat{\mathfrak{g}}_{W,b}(z_t)$ is obtained by CDF interpolation (see Appendix \ref{app:interpolation}) while the integral is computed numerically using the Gauss-Chebyshev quadrature formulas (\cite{judd1998}, section 7.2) with 300 Chebychev quadrature nodes on $[2y_{min},2y_{max}]$. CRPS score is of the highest value when the distributions are equal. We obtain an average CRPS score of the out-of-sample forecast as $CRPS_{OOS} = \frac{1}{T}\sum_{t=t_0+1}^{T} CRPS_t$. 
  
Another measure for the distributional forecast accuracy is Brier score \citep{gneiting2007strictly}. At time $t$, it calculates a squared difference of binary realization and the probability forecast,
\begin{equation}
B_t = - \sum_{j=1}^{p+1} \left(  \mathbb{I}_{\{q^{\alpha_{j-1}} < y_t \leq q^{\alpha_j}\}} - \widehat{\Pr} \{ q^{\alpha_{j-1}} < y_t \leq q^{\alpha_{j}} \} \right) ^2.
\end{equation}
We also compute the average value of the Brier score for the out-of-sample period.

We compare proposed models using relative predictive performance of two models, $M_1 / M_2$, where $M_i$ is particular measure (MPSE, CRPS, or Brier) corresponding model $i$.
The model $M_1$ performs better when the ratio is lower than one.

\subsection{Discussion}

Table~\ref{Tab:avg-overall-performance}, Figure~\ref{fig:benchScatter29assets3models} and Figure~\ref{fig:benchBoxplots29assets3models} in Appendix \ref{app:tables} provide all out-of-sample results with the horizon $h=1$ comparing sizes of various machine learning models and different variables used as predictors. The performance for all measures is put relative to the maximum likelihood ordered logit model, an AB benchmark of \cite{barunik2019simple}. 
\begin{table}[ht]
  \centering
  \begin{tabular}{rlllll}
    \toprule
    \textbf{Predictors} & Model & \textbf{Avg. MSPE}  & \textbf{Avg. CRPS} & \textbf{Avg. Brier score}\\\midrule
    AB & Ordered logit & 1.0 & 1.0 & 1.0 \\
    \cmidrule{2-5}
    \multirow{3}{*}{AB} & NN:128 & 1.00116 & 0.997133 & 1.00159 \\
     & NN:128x64 & 1.00147 & 0.999602 & 1.00069 \\
     & NN:128x64x32 & 1.00217 & 1.00317 & 1.00105 \\
     \cmidrule{2-5}
    \multirow{3}{*}{RM} & NN:128 & 0.991975  & 0.981237 & 0.995725 \\
     & NN:128x64 & 0.992387 & 0.981804 & 0.996069 \\
     & NN:128x64x32 & 0.992872 & 0.983318 & 0.996323 \\
     \cmidrule{2-5}
    \multirow{3}{*}{AB+RM} & NN:128 & 0.994755 &  0.98282 & 0.997473 \\
    & NN:128x64 & 0.99483  & 0.983535 & 0.997176 \\
     & NN:128x64x32 & 0.994522 & 0.984208 & 0.997309 \\
     \bottomrule
  \end{tabular}
  \caption{Performance results according to different scores, improvement}
\begin{minipage}{\textwidth} 
            \footnotesize
            The table shows performance of three models with different sizes and of three different input features. All results are bench-marked to Ordered Logit of \citet{barunik2019simple}. Value lower than 1 states that the purposed model is better than benchmark.
        \end{minipage} 
  \label{Tab:avg-overall-performance}
\end{table}

Overall, we document that machine learning is capable of forecasting conditional distribution of asset returns well, and providing informative variables it delivers improving predictions. Particularly, we observe an average 1\% out-of-sample improvements in MSPE for all studied assets in comparison to the parametric models. The improvement is even larger in terms of the Continuous rank probability score evaluating the compatibility of predicted and data distributions.

It is important to note here that in financial forecasting the relationship between statistical and economic gains from predictions is non-trivial. \cite{campbell2008predicting,Rapach:2010} note that seemingly small statistical improvement could generate large benefits in practice, which has recently been confirmed on expected returns forecast by machine learning \citep{gu2020empirical,babiak2020deep}. Hence an average 1~\% improvement in the out-of-sample predictions we document will most likely be interesting for a practitioners forming their portfolios based on our forecasts. While it is tempting to explore such strategy, it is far beyond the scope and space of this text.

\begin{figure}[ht!]
    \rotatebox{90}{\hspace{5em}MSPE}
    \includegraphics[width=0.98\textwidth]{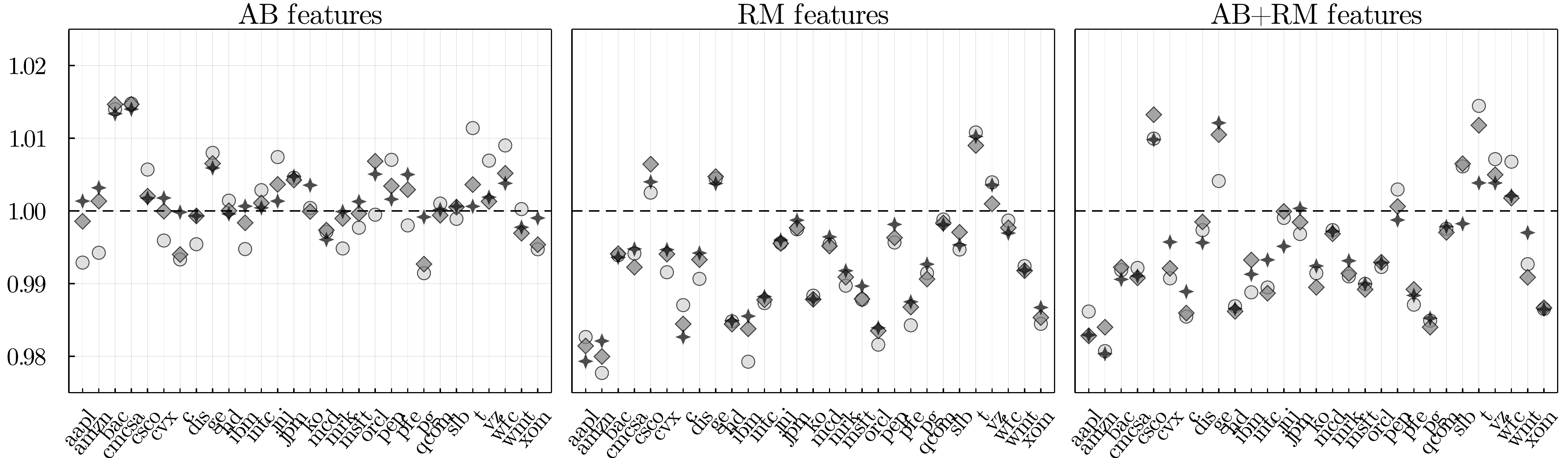}
    \rotatebox{90}{\hspace{5em}CRPS}
    \includegraphics[width=0.98\textwidth]{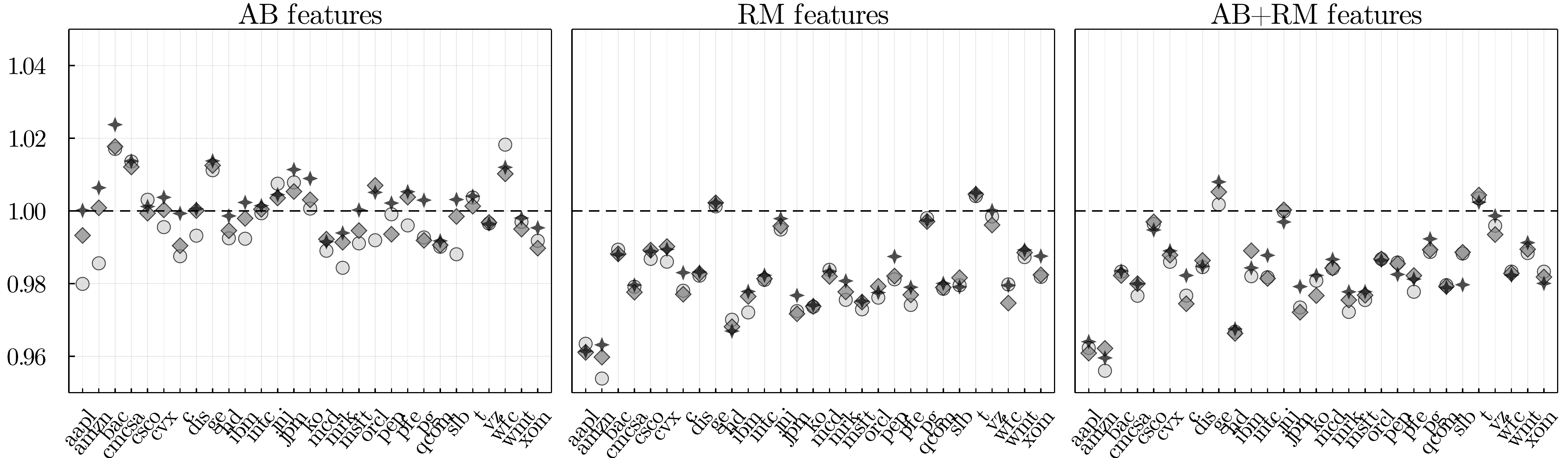}
    \rotatebox{90}{\hspace{5em}Brier}
    \includegraphics[width=0.98\textwidth]{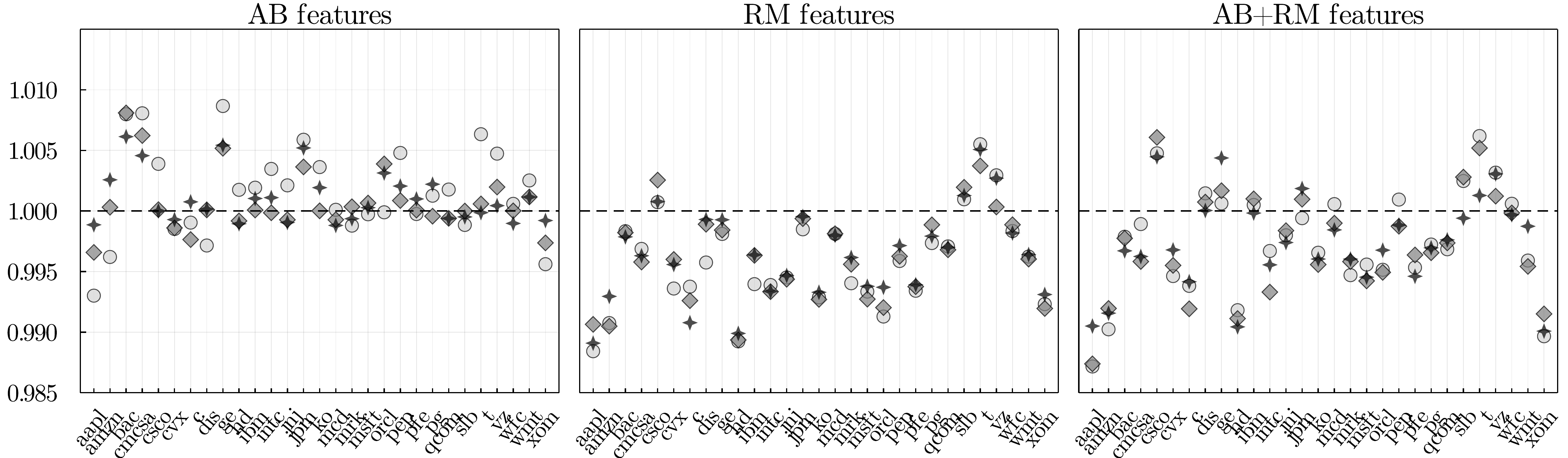}
    \caption{Comparison of the out-of-samples forecasts of 29 U.S. stocks}
\begin{minipage}{\textwidth} 
            \footnotesize
            The three statistical measures, MSPE (top), CRPS (middle), and Brier score (bottom) are used for the three set of predictors and three machine learning models depicted as $star$ for 128, $diamond$ for 128x64, and $circle$ for 128x64x32. \citet{barunik2019simple} ordered logit model is benchmark with value 1. Value lower than 1 show better better performance of a model in comparison to benchmark.
        \end{minipage} 
    \label{fig:benchScatter29assets3models}
\end{figure}

More specifically, Table~\ref{Tab:avg-overall-performance} shows that neural networks of all sizes on average bring approximately $\approx 0.1\%$ worse results in comparison to parametrized AB model when the same set of predictors is used and hence bring statistically equivalent forecasts. This result suggests that data does not contain any further non-linearities that are not captured by a parametric AB model, and since machine learning is more flexible in number of parameters to be estimated, it learns and approximates the AB parametrization with a small degree of error. With respect to CRSP, neural network models seem to be equivalent and hence the deep learning feature does not help in this case.

Situation changes with additional predictors when the AB approach becomes infeasible and machine learning approach offers possibility to explore how informative the predictors are for forecasts. When additional five realized measures (RM) are used as predictors, performance increases with respect to all measures. With respect to depth of networks, the shallow (NN 128) neural network shows best results. This result is similar to \citet{gu2020empirical} who find that shallow network performs better than deeper structures one on asset returns data. 

While the Table~\ref{Tab:avg-overall-performance} provides aggregate results for all stocks, Figures~\ref{fig:benchBoxplots29assets3models}~and~\ref{fig:benchScatter29assets3models} complement it with all measures reported for individual assets in boxplots. The detailed look uncovers that machine learning improves performance of individual stocks such as AAPL, AMZZN, GE, or WMT even more.
At the same time, Figure~\ref{fig:benchBoxplots29assets3models} shows that in most cases, deeper networks shows lower variance for most of stocks.

\section{Conclusion}
\label{sec: conclusion}

In this paper, we have proposed a new approach to modelling probability distributions of economic variables using the state-of-the-art machine learning methods. The distributional neural network relaxes the assumption on the distribution family of time series and lets the model to explore the data fully. The approach is particularly beneficial for modelling data with non-Gaussian, non-linear and asymmetric structures. We show the usefulness of the approach in an economic and financial application. At the same time teh approach is general and can be applied to any other dataset.

We have illustrated that our distributional neural network is useful in constructing big data-driven macroeconomic fan charts that are first of its kind since they are learned from the structure between 216 relevant economic variables. Further, we illustrate how deep-learning can be used to improve probabilistic forecasts of data notoriously know to contain low signal-to-noise ratio, heavy tails and asymmetries.

\begingroup
\linespread{1}
\setlength{\bibsep}{0pt}
\setlength{\bibhang}{1.0em}
\bibliographystyle{chicago}
\bibliography{BiblioDistributions}

\begin{thebibliography}{}

\bibitem[\protect\citeauthoryear{Anatolyev and Barun\'{i}k}{Anatolyev and
  Barun\'{i}k}{2019}]{barunik2019simple}
Anatolyev, S. and J.~Barun\'{i}k (2019).
\newblock Forecasting dynamic return distributions based on ordered binary
  choice.
\newblock {\em International Journal of Forecasting\/}~{\em 35\/}(3), 823--835.

\bibitem[\protect\citeauthoryear{Athey and Imbens}{Athey and
  Imbens}{2019}]{athey2019machine}
Athey, S. and G.~W. Imbens (2019).
\newblock Machine learning methods that economists should know about.
\newblock {\em Annual Review of Economics\/}~{\em 11}, 685--725.

\bibitem[\protect\citeauthoryear{Babiak and Barun{\'\i}k}{Babiak and
  Barun{\'\i}k}{2020}]{babiak2020deep}
Babiak, M. and J.~Barun{\'\i}k (2020).
\newblock Deep learning, predictability, and optimal portfolio returns.
\newblock {\em arXiv preprint arXiv:2009.03394\/}.

\bibitem[\protect\citeauthoryear{Baillie and Kapetanios}{Baillie and
  Kapetanios}{2007}]{baillie2007testing}
Baillie, R.~T. and G.~Kapetanios (2007).
\newblock Testing for neglected nonlinearity in long-memory models.
\newblock {\em Journal of Business \& Economic Statistics\/}~{\em 25\/}(4),
  447--461.

\bibitem[\protect\citeauthoryear{Bianchi, B{\"u}chner, and Tamoni}{Bianchi
  et~al.}{2020}]{bianchi2020bond}
Bianchi, D., M.~B{\"u}chner, and A.~Tamoni (2020).
\newblock Bond risk premia with machine learning.
\newblock {\em Review of Financial Studies\/}~(forthcoming).

\bibitem[\protect\citeauthoryear{Box, Draper, et~al.}{Box
  et~al.}{1987}]{box1987empirical}
Box, G.~E., N.~R. Draper, et~al. (1987).
\newblock {\em Empirical model-building and response surfaces}, Volume 424.
\newblock Wiley New York.

\bibitem[\protect\citeauthoryear{Box, Jenkins, Reinsel, and Ljung}{Box
  et~al.}{2015}]{box2015time}
Box, G.~E., G.~M. Jenkins, G.~C. Reinsel, and G.~M. Ljung (2015).
\newblock {\em Time series analysis: forecasting and control}.
\newblock John Wiley \& Sons.

\bibitem[\protect\citeauthoryear{Breiman et~al.}{Breiman
  et~al.}{2001}]{breiman2001statistical}
Breiman, L. et~al. (2001).
\newblock Statistical modeling: The two cultures (with comments and a rejoinder
  by the author).
\newblock {\em Statistical science\/}~{\em 16\/}(3), 199--231.

\bibitem[\protect\citeauthoryear{Britton, Fisher, and Whitley}{Britton
  et~al.}{1998}]{britton1998inflation}
Britton, E., P.~Fisher, and J.~Whitley (1998).
\newblock The inflation report projections: understanding the fan chart.
\newblock {\em Chart\/}~{\em 8\/}(10).

\bibitem[\protect\citeauthoryear{Bryzgalova, Pelger, and Zhu}{Bryzgalova
  et~al.}{2019}]{bryzgalova2019forest}
Bryzgalova, S., M.~Pelger, and J.~Zhu (2019).
\newblock Forest through the trees: Building cross-sections of stock returns.
\newblock {\em Available at SSRN 3493458\/}.

\bibitem[\protect\citeauthoryear{Campbell and Thompson}{Campbell and
  Thompson}{2008}]{campbell2008predicting}
Campbell, J.~Y. and S.~B. Thompson (2008).
\newblock Predicting excess stock returns out of sample: Can anything beat the
  historical average?
\newblock {\em The Review of Financial Studies\/}~{\em 21\/}(4), 1509--1531.

\bibitem[\protect\citeauthoryear{Chen, Pelger, and Zhu}{Chen
  et~al.}{2020}]{chen2020deep}
Chen, L., M.~Pelger, and J.~Zhu (2020).
\newblock Deep learning in asset pricing.
\newblock {\em Available at SSRN 3350138\/}.

\bibitem[\protect\citeauthoryear{Chen, Kang, Chen, and Wang}{Chen
  et~al.}{2020}]{chen2020probabilistic}
Chen, Y., Y.~Kang, Y.~Chen, and Z.~Wang (2020).
\newblock Probabilistic forecasting with temporal convolutional neural network.
\newblock {\em Neurocomputing\/}.

\bibitem[\protect\citeauthoryear{Chernozhukov, Fern{\'a}ndez-Val, and
  Melly}{Chernozhukov et~al.}{2013}]{chernozhukov2013inference}
Chernozhukov, V., I.~Fern{\'a}ndez-Val, and B.~Melly (2013).
\newblock Inference on counterfactual distributions.
\newblock {\em Econometrica\/}~{\em 81\/}(6), 2205--2268.

\bibitem[\protect\citeauthoryear{Clements, Galv{\~a}o, and Kim}{Clements
  et~al.}{2008}]{clements2008quantile}
Clements, M.~P., A.~B. Galv{\~a}o, and J.~H. Kim (2008).
\newblock Quantile forecasts of daily exchange rate returns from forecasts of
  realized volatility.
\newblock {\em Journal of Empirical Finance\/}~{\em 15\/}(4), 729--750.

\bibitem[\protect\citeauthoryear{Coulombe, Leroux, Stevanovic, and
  Surprenant}{Coulombe et~al.}{2020}]{coulombe2020machine}
Coulombe, P.~G., M.~Leroux, D.~Stevanovic, and S.~Surprenant (2020).
\newblock How is machine learning useful for macroeconomic forecasting?
\newblock {\em arXiv preprint arXiv:2008.12477\/}.

\bibitem[\protect\citeauthoryear{Diebold}{Diebold}{2020}]{diebold2020big}
Diebold, F.~X. (2020).
\newblock " big data" and its origins.
\newblock {\em arXiv preprint arXiv:2008.05835\/}.

\bibitem[\protect\citeauthoryear{Diebold and Mariano}{Diebold and
  Mariano}{1995}]{diebold1995comparing}
Diebold, F.~X. and R.~S. Mariano (1995).
\newblock Comparing predictive accuracy.
\newblock {\em Journal of Business \& Economic Statistics\/}~{\em 13\/}(3).

\bibitem[\protect\citeauthoryear{Duan, Avati, Ding, Basu, Ng, and Schuler}{Duan
  et~al.}{2019}]{duan2019ngboost}
Duan, T., A.~Avati, D.~Y. Ding, S.~Basu, A.~Y. Ng, and A.~Schuler (2019).
\newblock Ngboost: Natural gradient boosting for probabilistic prediction.
\newblock {\em arXiv preprint arXiv:1910.03225\/}.

\bibitem[\protect\citeauthoryear{Engle and Manganelli}{Engle and
  Manganelli}{2004}]{engle2004caviar}
Engle, R.~F. and S.~Manganelli (2004).
\newblock Caviar: Conditional autoregressive value at risk by regression
  quantiles.
\newblock {\em Journal of Business \& Economic Statistics\/}~{\em 22\/}(4),
  367--381.

\bibitem[\protect\citeauthoryear{Fama}{Fama}{1965}]{fama1965portfolio}
Fama, E.~F. (1965).
\newblock Portfolio analysis in a stable paretian market.
\newblock {\em Management science\/}~{\em 11\/}(3), 404--419.

\bibitem[\protect\citeauthoryear{Feng, He, and Polson}{Feng
  et~al.}{2018}]{feng2018deep}
Feng, G., J.~He, and N.~G. Polson (2018).
\newblock Deep learning for predicting asset returns.
\newblock {\em arXiv preprint arXiv:1804.09314\/}.

\bibitem[\protect\citeauthoryear{Foresi and Peracchi}{Foresi and
  Peracchi}{1995}]{foresi1995conditional}
Foresi, S. and F.~Peracchi (1995).
\newblock The conditional distribution of excess returns: An empirical
  analysis.
\newblock {\em Journal of the American Statistical Association\/}~{\em
  90\/}(430), 451--466.

\bibitem[\protect\citeauthoryear{Fortin, Lemieux, and Firpo}{Fortin
  et~al.}{2011}]{fortin2011decomposition}
Fortin, N., T.~Lemieux, and S.~Firpo (2011).
\newblock Decomposition methods in economics.
\newblock In {\em Handbook of labor economics}, Volume~4, pp.\  1--102.
  Elsevier.

\bibitem[\protect\citeauthoryear{Friedman}{Friedman}{2009}]{friedman2009harvard}
Friedman, W.~A. (2009).
\newblock The harvard economic service and the problems of forecasting.
\newblock {\em History of Political Economy\/}~{\em 41\/}(1), 57--88.

\bibitem[\protect\citeauthoryear{Fritsch and Carlson}{Fritsch and
  Carlson}{1980}]{fritsch1980interpolation}
Fritsch, F. and R.~Carlson (1980).
\newblock Monotone piecewise cubic interpolation.
\newblock {\em SIAM Journal on Numerical Analysis\/}~{\em 17\/}(2).

\bibitem[\protect\citeauthoryear{Gasthaus, Benidis, Wang, Rangapuram, Salinas,
  Flunkert, and Januschowski}{Gasthaus
  et~al.}{2019}]{gasthaus2019probabilistic}
Gasthaus, J., K.~Benidis, Y.~Wang, S.~S. Rangapuram, D.~Salinas, V.~Flunkert,
  and T.~Januschowski (2019).
\newblock Probabilistic forecasting with spline quantile function rnns.
\newblock In {\em The 22nd International Conference on Artificial Intelligence
  and Statistics}, pp.\  1901--1910.

\bibitem[\protect\citeauthoryear{Geweke and Whiteman}{Geweke and
  Whiteman}{2006}]{geweke2006bayesian}
Geweke, J. and C.~Whiteman (2006).
\newblock Bayesian forecasting.
\newblock {\em Handbook of economic forecasting\/}~{\em 1}, 3--80.

\bibitem[\protect\citeauthoryear{Gneiting}{Gneiting}{2008}]{gneiting2008probabilistic}
Gneiting, T. (2008).
\newblock Probabilistic forecasting.
\newblock {\em Journal of the Royal Statistical Society. Series A (Statistics
  in Society)\/}, 319--321.

\bibitem[\protect\citeauthoryear{Gneiting and Raftery}{Gneiting and
  Raftery}{2007}]{gneiting2007strictly}
Gneiting, T. and A.~E. Raftery (2007).
\newblock Strictly proper scoring rules, prediction, and estimation.
\newblock {\em Journal of the American statistical Association\/}~{\em
  102\/}(477), 359--378.

\bibitem[\protect\citeauthoryear{Gu, Kelly, and Xiu}{Gu
  et~al.}{2020}]{gu2020empirical}
Gu, S., B.~Kelly, and D.~Xiu (2020).
\newblock Empirical asset pricing via machine learning.
\newblock {\em The Review of Financial Studies\/}~{\em 33\/}(5), 2223--2273.

\bibitem[\protect\citeauthoryear{Heaton, Polson, and Witte}{Heaton
  et~al.}{2017}]{heaton2017deep}
Heaton, J.~B., N.~G. Polson, and J.~H. Witte (2017).
\newblock Deep learning for finance: deep portfolios.
\newblock {\em Applied Stochastic Models in Business and Industry\/}~{\em
  33\/}(1), 3--12.

\bibitem[\protect\citeauthoryear{Hersbach}{Hersbach}{2000}]{hersbach2000decomposition}
Hersbach, H. (2000).
\newblock Decomposition of the continuous ranked probability score for ensemble
  prediction systems.
\newblock {\em Weather and Forecasting\/}~{\em 15\/}(5), 559--570.

\bibitem[\protect\citeauthoryear{Hochreiter and Schmidhuber}{Hochreiter and
  Schmidhuber}{1997}]{hochreiter1997long}
Hochreiter, S. and J.~Schmidhuber (1997).
\newblock Long short-term memory.
\newblock {\em Neural computation\/}~{\em 9\/}(8), 1735--1780.

\bibitem[\protect\citeauthoryear{Hu, Guo, Li, Shen, and Sun}{Hu
  et~al.}{2019}]{hu2019distribution}
Hu, T., Q.~Guo, Z.~Li, X.~Shen, and H.~Sun (2019).
\newblock Distribution-free probability density forecast through deep neural
  networks.
\newblock {\em IEEE Transactions on Neural Networks and Learning
  Systems\/}~{\em 31\/}(2), 612--625.

\bibitem[\protect\citeauthoryear{Hutchinson, Lo, and Poggio}{Hutchinson
  et~al.}{1994}]{hutchinson1994nonparametric}
Hutchinson, J.~M., A.~W. Lo, and T.~Poggio (1994).
\newblock A nonparametric approach to pricing and hedging derivative securities
  via learning networks.
\newblock {\em The Journal of Finance\/}~{\em 49\/}(3), 851--889.

\bibitem[\protect\citeauthoryear{Hyndman, Koehler, Ord, and Snyder}{Hyndman
  et~al.}{2008}]{hyndman2008forecasting}
Hyndman, R., A.~B. Koehler, J.~K. Ord, and R.~D. Snyder (2008).
\newblock {\em Forecasting with exponential smoothing: the state space
  approach}.
\newblock Springer Science \& Business Media.

\bibitem[\protect\citeauthoryear{Innes, Saba, Fischer, Gandhi, Rudilosso, Joy,
  Karmali, Pal, and Shah}{Innes et~al.}{2018}]{Flux2018}
Innes, M., E.~Saba, K.~Fischer, D.~Gandhi, M.~C. Rudilosso, N.~M. Joy,
  T.~Karmali, A.~Pal, and V.~Shah (2018).
\newblock Fashionable modelling with flux.
\newblock {\em CoRR\/}~{\em abs/1811.01457}.

\bibitem[\protect\citeauthoryear{Israel, Kelly, and Moskowitz}{Israel
  et~al.}{2020}]{israel2020can}
Israel, R., B.~T. Kelly, and T.~J. Moskowitz (2020).
\newblock Can machines' learn'finance?
\newblock {\em Available at SSRN 3624052\/}.

\bibitem[\protect\citeauthoryear{Iworiso and Vrontos}{Iworiso and
  Vrontos}{2020}]{iworiso2020directional}
Iworiso, J. and S.~Vrontos (2020).
\newblock On the directional predictability of equity premium using machine
  learning techniques.
\newblock {\em Journal of Forecasting\/}~{\em 39\/}(3), 449--469.

\bibitem[\protect\citeauthoryear{Januschowski, Gasthaus, Wang, Salinas,
  Flunkert, Bohlke-Schneider, and Callot}{Januschowski
  et~al.}{2020}]{januschowski2020criteria}
Januschowski, T., J.~Gasthaus, Y.~Wang, D.~Salinas, V.~Flunkert,
  M.~Bohlke-Schneider, and L.~Callot (2020).
\newblock Criteria for classifying forecasting methods.
\newblock {\em International Journal of Forecasting\/}~{\em 36\/}(1), 167--177.

\bibitem[\protect\citeauthoryear{Judd}{Judd}{1998}]{judd1998}
Judd, K. (1998).
\newblock {\em Numerical Methods in Economics}.
\newblock Cambridge, MA: MIT Press.

\bibitem[\protect\citeauthoryear{Kingma and Ba}{Kingma and
  Ba}{2014}]{kingma2014adam}
Kingma, D.~P. and J.~Ba (2014).
\newblock Adam: A method for stochastic optimization.
\newblock {\em arXiv preprint arXiv:1412.6980\/}.

\bibitem[\protect\citeauthoryear{Koenker and Bassett~Jr}{Koenker and
  Bassett~Jr}{1978}]{koenker1978regression}
Koenker, R. and G.~Bassett~Jr (1978).
\newblock Regression quantiles.
\newblock {\em Econometrica: journal of the Econometric Society\/}, 33--50.

\bibitem[\protect\citeauthoryear{Kuan and White}{Kuan and
  White}{1994}]{kuan1994artificial}
Kuan, C.-M. and H.~White (1994).
\newblock Artificial neural networks: An econometric perspective.
\newblock {\em Econometric reviews\/}~{\em 13\/}(1), 1--91.

\bibitem[\protect\citeauthoryear{Lahiri, Martin, et~al.}{Lahiri
  et~al.}{2010}]{lahiri2010bayesian}
Lahiri, K., G.~Martin, et~al. (2010).
\newblock Bayesian forecasting in economics.
\newblock {\em International Journal of Forecasting\/}~{\em 26\/}(2), 211--215.

\bibitem[\protect\citeauthoryear{Leorato and Peracchi}{Leorato and
  Peracchi}{2015}]{leorato2015comparing}
Leorato, S. and F.~Peracchi (2015).
\newblock Comparing distribution and quantile regression.

\bibitem[\protect\citeauthoryear{Lerner}{Lerner}{1947}]{lerner1947measuring}
Lerner, A.~P. (1947).
\newblock Measuring business cycles. by arthur f. burns and wesley c.
  mitchell.[studies in business cycles, vol. ii.] new york: National bureau of
  economic research, 1946. pp. xxvii, 560.
\newblock {\em The Journal of Economic History\/}~{\em 7\/}(2), 222--226.

\bibitem[\protect\citeauthoryear{Lim and Gorse}{Lim and
  Gorse}{2020}]{lim2020deep}
Lim, Y.-S. and D.~Gorse (2020).
\newblock Deep probabilistic modelling of price movements for high-frequency
  trading.
\newblock {\em arXiv preprint arXiv:2004.01498\/}.

\bibitem[\protect\citeauthoryear{Lopez~de Prado}{Lopez~de
  Prado}{2019}]{lopez2019beyond}
Lopez~de Prado, M. (2019).
\newblock Beyond econometrics: A roadmap towards financial machine learning.
\newblock {\em Available at SSRN 3365282\/}.

\bibitem[\protect\citeauthoryear{Loshchilov and Hutter}{Loshchilov and
  Hutter}{2019}]{loshchilov2019adamw}
Loshchilov, I. and F.~Hutter (2019).
\newblock Decoupled weight decay regularization.
\newblock {\em arXiv preprint arXiv:1711.05101\/}.

\bibitem[\protect\citeauthoryear{Matheson and Winkler}{Matheson and
  Winkler}{1976}]{matheson1976scoring}
Matheson, J.~E. and R.~L. Winkler (1976).
\newblock Scoring rules for continuous probability distributions.
\newblock {\em Management science\/}~{\em 22\/}(10), 1087--1096.

\bibitem[\protect\citeauthoryear{McCracken and Ng}{McCracken and
  Ng}{2020}]{mccracken2020fred}
McCracken, M. and S.~Ng (2020).
\newblock Fred-qd: A quarterly database for macroeconomic research.
\newblock Technical report, National Bureau of Economic Research.

\bibitem[\protect\citeauthoryear{Mullainathan and Spiess}{Mullainathan and
  Spiess}{2017}]{mullainathan2017machine}
Mullainathan, S. and J.~Spiess (2017).
\newblock Machine learning: an applied econometric approach.
\newblock {\em Journal of Economic Perspectives\/}~{\em 31\/}(2), 87--106.

\bibitem[\protect\citeauthoryear{Racine}{Racine}{2001}]{racine2001nonlinear}
Racine, J. (2001).
\newblock On the nonlinear predictability of stock returns using financial and
  economic variables.
\newblock {\em Journal of Business \& Economic Statistics\/}~{\em 19\/}(3),
  380--382.

\bibitem[\protect\citeauthoryear{Rapach, Strauss, and Zhou}{Rapach
  et~al.}{2010}]{Rapach:2010}
Rapach, D., J.~Strauss, and G.~Zhou (2010).
\newblock Out-of-sample equity premium prediction: Combination forecasts and
  links to the real economy.
\newblock {\em Review of Financial Studies\/}~{\em 23\/}(2), 821--862.

\bibitem[\protect\citeauthoryear{Rothe}{Rothe}{2012}]{rothe2012partial}
Rothe, C. (2012).
\newblock Partial distributional policy effects.
\newblock {\em Econometrica\/}~{\em 80\/}(5), 2269--2301.

\bibitem[\protect\citeauthoryear{Salinas, Flunkert, Gasthaus, and
  Januschowski}{Salinas et~al.}{2020}]{salinas2020deepar}
Salinas, D., V.~Flunkert, J.~Gasthaus, and T.~Januschowski (2020).
\newblock Deepar: Probabilistic forecasting with autoregressive recurrent
  networks.
\newblock {\em International Journal of Forecasting\/}~{\em 36\/}(3),
  1181--1191.

\bibitem[\protect\citeauthoryear{Sirignano, Sadhwani, and Giesecke}{Sirignano
  et~al.}{2016}]{sirignano2016deep}
Sirignano, J., A.~Sadhwani, and K.~Giesecke (2016).
\newblock Deep learning for mortgage risk.
\newblock {\em arXiv preprint arXiv:1607.02470\/}.

\bibitem[\protect\citeauthoryear{Srivastava, Hinton, Krizhevsky, Sutskever, and
  Salakhutdinov}{Srivastava et~al.}{2014}]{srivastava2014dropout}
Srivastava, N., G.~Hinton, A.~Krizhevsky, I.~Sutskever, and R.~Salakhutdinov
  (2014).
\newblock Dropout: a simple way to prevent neural networks from overfitting.
\newblock {\em The Journal of Machine Learning Research\/}~{\em 15\/}(1),
  1929--1958.

\bibitem[\protect\citeauthoryear{Stock and Watson}{Stock and
  Watson}{2017}]{stock2017twenty}
Stock, J.~H. and M.~W. Watson (2017).
\newblock Twenty years of time series econometrics in ten pictures.
\newblock {\em Journal of Economic Perspectives\/}~{\em 31\/}(2), 59--86.

\bibitem[\protect\citeauthoryear{Timmermann}{Timmermann}{2000}]{timmermann2000density}
Timmermann, A. (2000).
\newblock Density forecasting in economics and finance.
\newblock {\em Journal of Forecasting\/}~{\em 19\/}(4), 231.

\bibitem[\protect\citeauthoryear{Tobek and Hronec}{Tobek and
  Hronec}{2020}]{tobek2020does}
Tobek, O. and M.~Hronec (2020).
\newblock Does it pay to follow anomalies research? machine learning approach
  with international evidence.
\newblock {\em Journal of Financial Markets\/}, 100588.

\bibitem[\protect\citeauthoryear{Wen, Torkkola, Narayanaswamy, and Madeka}{Wen
  et~al.}{2017}]{wen2017multi}
Wen, R., K.~Torkkola, B.~Narayanaswamy, and D.~Madeka (2017).
\newblock A multi-horizon quantile recurrent forecaster.
\newblock {\em arXiv preprint arXiv:1711.11053\/}.

\bibitem[\protect\citeauthoryear{Zhang, Zohren, and Roberts}{Zhang
  et~al.}{2020}]{zhang2020deep}
Zhang, Z., S.~Zohren, and S.~Roberts (2020).
\newblock Deep learning for portfolio optimisation.
\newblock {\em arXiv preprint arXiv:2005.13665\/}.

\bibitem[\protect\citeauthoryear{{\v{Z}}ike{\v{s}} and
  Barun\'{i}k}{{\v{Z}}ike{\v{s}} and Barun\'{i}k}{2016}]{vzikevs2014semi}
{\v{Z}}ike{\v{s}}, F. and J.~Barun\'{i}k (2016).
\newblock Semi-parametric conditional quantile models for financial returns and
  realized volatility.
\newblock {\em Journal of Financial Econometrics\/}~{\em 14\/}(1), 185--226.

\end{thebibliography}
\endgroup

\newpage
\setcounter{section}{0}
\setcounter{equation}{0}
\setcounter{figure}{0}
\setcounter{table}{0}

\def\thesection{\Alph{section}}
\def\thesubsection{\thesection.\arabic{subsection}}
\def\thesubsubsection{\thesubsection.\arabic{subsubsection}}
\renewcommand{\theequation}{\Alph{section}.\arabic{equation}}
\renewcommand{\thetable}{A\arabic{table}}
\renewcommand{\thefigure}{A\arabic{figure}}

\begin{center}
    \Large \textbf{Appendix for} 
\end{center}
\begin{center}
    \Large
    ``Learning Probability Distributions of Economic Variables''
\end{center}


\section{CDF interpolation}
\label{app:interpolation}
The Fritsch--Carlson monotonic cubic interpolation \citep{fritsch1980interpolation}
provides a monotonically increasing CDF with range $[0,1]$ when applied to CDF estimates on a finite grid.

Suppose we have CDF $F(y)$ defined at points $(y_k,F(y_k))$ for $k=1,\dots ,K,$ where $F(y_0)=0$ and $F(y_K)=1$.
We presume that $y_k<y_{k+1}$ and $F(y_k)<F(y_{k+1})$ for all $k=0,\dots ,K-1,$ which is warranted by continuity of returns and construction of the estimated distribution. First, we compute slopes of the secant lines as $\Delta_k =(F(y_{k+1})-F(y_k)))/(y_{k+1}-y_k)$ for $k=1,\dots,K-1,$ and then the tangents at every data point as $m_1 = \Delta_1$, $m_k = \frac{1}{2}(\Delta_{k-1}+\Delta_k)$ for $k=2,\dots,K-1$, and $m_K = \Delta_{K-1}.$
Let $\alpha_k = m_k/\Delta_k$ and $\beta_k = m_{k+1}/\Delta_k$ for $k=1,\dots,K-1$.
If $\alpha_k^2 + \beta_k^2 > 9$ for some $k=1,\dots,K-1,$ then we set $m_k = \tau_k \alpha_k \Delta_k$ and $m_{k+1} = \tau_k \beta_k \Delta_k,$ with $\tau_k = 3(\alpha_k^2 + \beta_k^2)^{-1/2}$.
Finally, the cubic Hermite spline is applied: for any $y\in [y_k,y_{k+1}]$ for some $k=0,\dots,K-1,$
we evaluate $F(y)$ as
$$F(y) = (2t^3-3t^2+1)F(y_k)  + (t^3-2t^2+t)h y_{k}  + (-2t^3+3t^2) F(y_{k+1}) + (t^3-t^2)h m_{k+1},$$ where $h=y_{k+1}-y_{k}$ and $t=(y-y_{k})/h.$

\clearpage
\section{Additional tables and figures}
\label{app:tables}

\begin{table}[ht!]
  \centering
  \begin{tabular}{llll}
    \toprule
    \textbf{Hyper parameters} & \textbf{Values} & \textbf{} \\\midrule
    Learning rate, $\eta$ &  0.0001, 0.001, 0.005 & \\
    Dropout rate, $\phi$ & 0.2, 0.4 &   \\
    $L_2$-decay regularization rate, $\lambda_W$ & 0.00001, 0.00005 &   \\
    Nodes dimensions & 32x32, 64x64, 60x50 & & \\\toprule
    \textbf{Fixed parameters} & Value & \\\midrule
    Number of layers & 2 & \\
    Mini batch size & 8 & \\
    Epochs & 350 & \\
    Monotonicity parameter, $\lambda_m$ & 5.0 & \\
    Cross-validation, k-folds & 3 & \\
    Train/test ratio & 0.93 & \\
    Ensembles & 1 & \\\bottomrule
  \end{tabular}
  \caption{Fan chart recurrent DistrNN parameters space for the empirical application, Sec.~ \ref{sec: empirics-macro}}
  	\begin{minipage}{\textwidth} 
	    \footnotesize
	    The hyperoptimization algorithm searches through the whole hyperparameter space and tries all sets/combinations of hyperparameters to evaluate the model.
    \end{minipage} 
  \label{Tab:pars-macro}
\end{table}

\begin{table}[ht]
  \centering
  \begin{tabular}{llll}
    \toprule
    \textbf{Hyper parameters} & \textbf{Minimum value} & \textbf{Maximum value} \\\midrule
    Learning rate, $\eta$ & 0.0001 &  0.02 \\
    Dropout rate, $\phi$ & 0.0 &  0.5 \\
    $L_2$-decay regularization rate, $\lambda_W$ & 0.0 &  0.0018 \\\toprule
    \textbf{Fixed parameters} & Value & \\\midrule
    Epochs & 250 & \\
    Early stopping patience & 25 & \\
    Monotonicity parameter, $\lambda_m$ & 0.2 & \\
    Mini batch size & 32 & \\
    Ensembles & 3 & \\
    Number of layers & 1, 2, 3 & \\
    Nodes dimensions & 128, 128x64, 128x64x32 & \\\bottomrule
  \end{tabular}
  \caption{DistrNN parameters space for the emprical application, Sec.~\ref{sec: results-assets}}
    \begin{minipage}{\textwidth} 
	    \footnotesize
	    The hyperoptimization algorithm searches through the hyperparameter space and randomly tries sets of parameters to evaluate the model.
    \end{minipage} 
  \label{Tab:pars-assets}
\end{table} 

\begin{figure}[ht!]
    \centering
    \rotatebox{90}{\hspace{5em}MSPE}\includegraphics[width=0.98\textwidth]{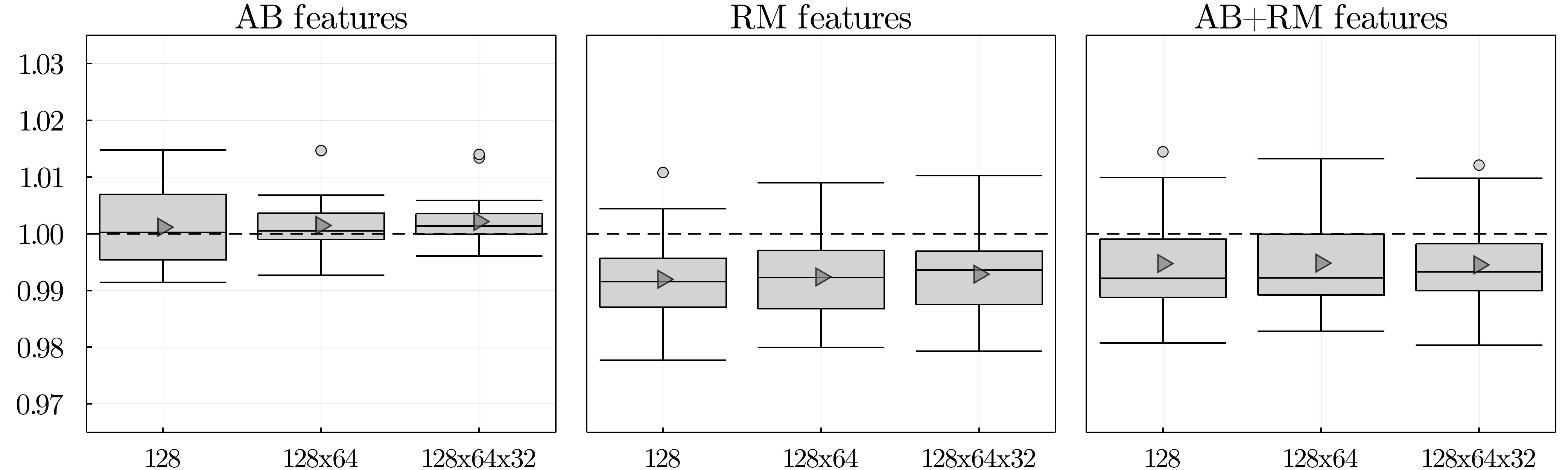}\\[0.1em]
    \rotatebox{90}{\hspace{5em}CRPS}\includegraphics[width=0.98\textwidth]{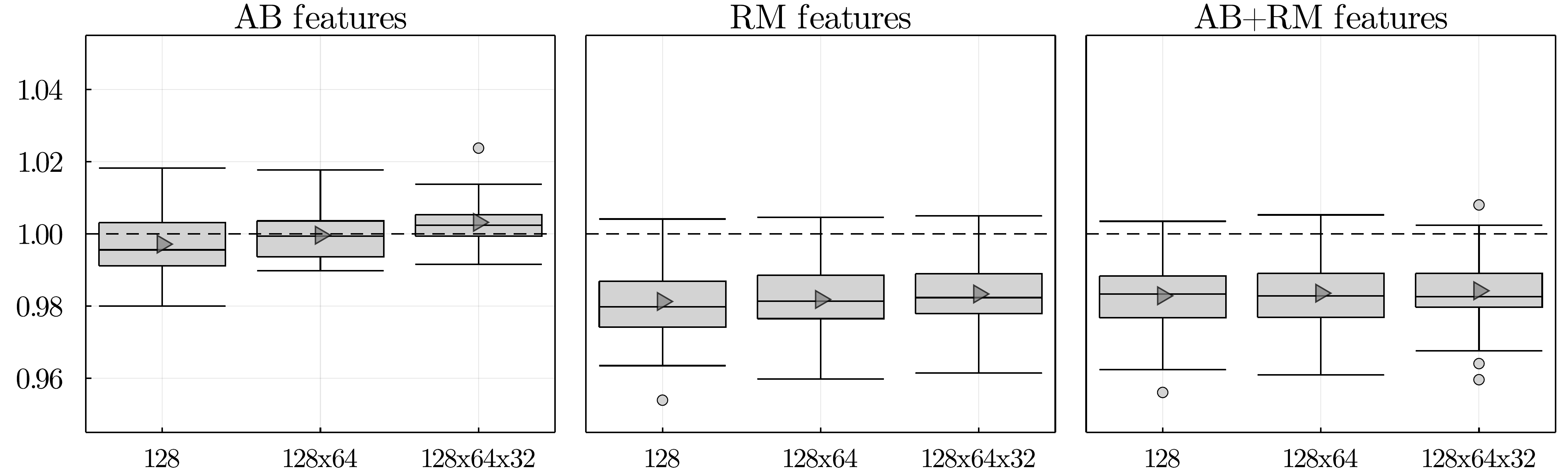}\\[0.1em]
    \rotatebox{90}{\hspace{5em}Brier}\includegraphics[width=0.98\textwidth]{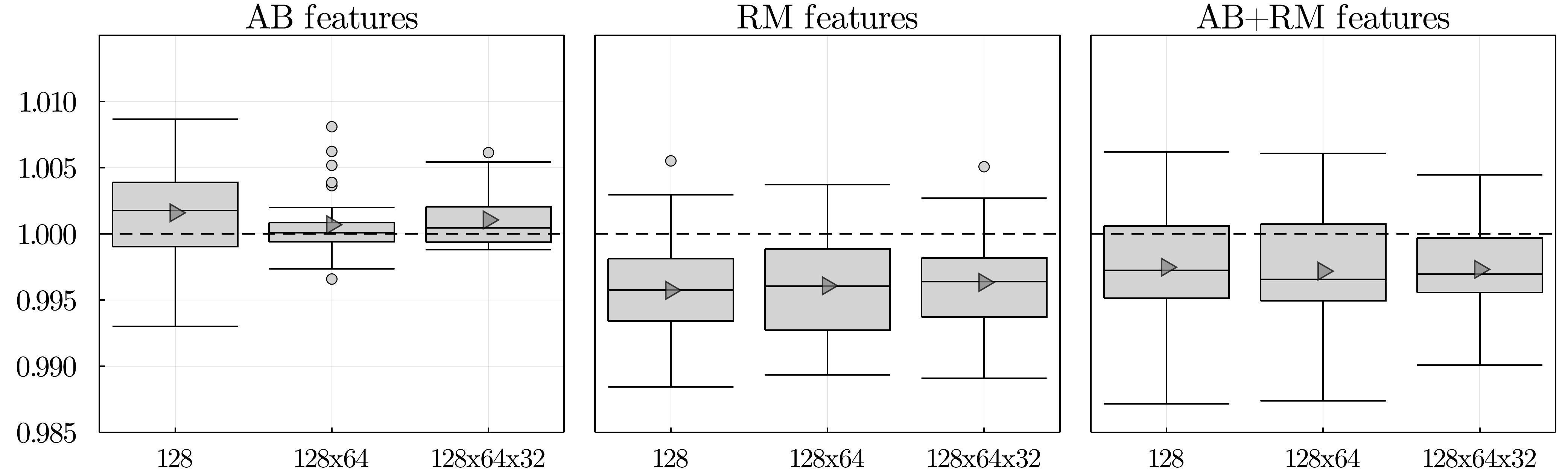}
    \caption{Comparison of the out-of-sample forecasts.}
    \begin{minipage}{\textwidth} 
            \footnotesize
            The three statistical measures: MSPE (top), CRPS (middle), and Brier (bottom). Each box-plot depicts benchmark values of 29 assets of given NN model size. \citet{barunik2019simple} ordered logit model is benchmark=1. Value lower than 1 states that the purposed model is better that benchmark.
        \end{minipage} 

    \label{fig:benchBoxplots29assets3models}
\end{figure}

\begin{table}[ht]
  \centering
  \begin{tabular}{llrrrr}
    \toprule
    \textbf{Data} & Model & \textbf{MSPE} & \textbf{Bin.CE} & \textbf{CRPS} & \textbf{Brier} \\\midrule
    AB & NN:128 & 14/29 & 14/29 & 20/29 & 11/29 \\
    AB & NN:128x64 & 13/29 & 11/29 & 15/29 & 11/29 \\
    AB & NN:128x64x32 & 8/29 & 7/29 & 8/29 & 11/29 \\\midrule
    RM & NN:128 & 25/29 & 25/29 & 27/29 & 25/29 \\
    RM & NN:128x64 & 25/29 & 24/29 & 27/29 & 25/29 \\
    RM & NN:128x64x32 & 25/29 & 24/29 & 26/29 & 25/29 \\\midrule
    AB+RM & NN:128 & 22/29 & 19/29 & 27/29 & 19/29 \\
    AB+RM & NN:128x64 & 22/29 & 21/29 & 26/29 & 21/29 \\
    AB+RM & NN:128x64x32 & 23/29 & 23/29 & 27/29 & 23/29 \\\bottomrule
  \end{tabular}
  \caption{Results according to different scores, assets}
  \label{Tab: assets-count-better}
  \begin{minipage}{\textwidth} 
    	\footnotesize
		The table shows performance of three models with different sizes and of three different input features. All results are benchmarked to Ordered Logit of \citet{barunik2019simple}. The number indicates for how many assets given model performs better.
  \end{minipage} 
\end{table}

\end{document}